\documentclass{article}

\usepackage[usenames,dvipsnames,svgnames,x11names]{xcolor}
\newcommand{\ysnote}[1]{ {\textcolor{magenta} { ***Yogesh: #1 }}} 
\newcommand{\drnote}[1]{ {\textcolor{orange} { ***Dreamer: #1 }}}
\newcommand{\Note}[1]{\textcolor{red}{#1}} 
\newcommand{\ysnoted}[1]{ {\textcolor{green} { ***TODO Later: #1 }}} 
\renewcommand{\ysnoted}[1]{} 

\usepackage[T1]{fontenc} 
\usepackage{lmodern} 
\usepackage[normalem]{ulem} 

\usepackage{blindtext}

\usepackage{multirow}
\usepackage{graphicx,epstopdf}
\usepackage{comment}
\usepackage{color}
\usepackage[hyphens]{url}
\usepackage{hyperref}
\usepackage{amsmath}
\usepackage{amssymb}
\usepackage{cite}
\usepackage{array}
\usepackage{soul}
\usepackage{pdflscape}
\usepackage{afterpage}
\usepackage{rotating}
\usepackage{caption}
\usepackage{subfig}
\usepackage{float}
\usepackage[utf8]{inputenc}
\usepackage[english]{babel}
\usepackage{longtable}
\usepackage{supertabular,booktabs}
\usepackage[font=small,labelfont=bf]{caption}
\usepackage{wrapfig}
\usepackage{authblk}
\usepackage[T1]{fontenc}
\usepackage{lipsum} 
\usepackage{pdfpages}
\usepackage[graphicx]{realboxes}
\usepackage{longtable}
\usepackage{multirow}
\usepackage{float}
\usepackage{tabularx}
\usepackage{csquotes} 

\usepackage[nopostdot,nonumberlist]{glossaries}
\makeglossaries
\newglossaryentry{latex}
{
    name=latex,
    description={Is a mark up language specially suited 
    for scientific documents}
}

\newglossaryentry{maths}
{
    name=mathematics,
    description={Mathematics is what mathematicians do}
}

\newglossaryentry{SocialDistancing}
{
    name={Social Distancing},
    description={Social Distancing is the practice of maintaining physical distance between individuals to prevent the spread of face-to-face communicable diseases. A $1.5$--$2~m$ distance is recommended for COVID-19.}
}

\newglossaryentry{ContactTracing}
{
    name={Contact Tracing},
    description={Contact Tracing is the process of identifying people might be at risk due to physical interactions with a disease carrier.}
}

\newglossaryentry{DCT}
{
    name={Digital Contact Tracing},
    description={Digital Contact Tracing refers to the use of technology, such as mobile phone apps, QR codes, wearable devices and video analytics, to assist with contact tracing.}
}

\newglossaryentry{Bluetooth}
{
    name={Bluetooth},
    description={Bluetooth is a wireless technology standard for short-range communication between mobile devices such as laptops and smart phones, with a practical range of up to $10~m$ meters.}
}

\newglossaryentry{BLE}
{
    name={Bluetooth Low Energy (BLE)},
    description={Bluetooth Low Energy (BLE) is a variant of the Bluetooth standard which uses much lesser power for communication, allowing it to be enabled all the time. It is widely used in smart phones, wearables, beacons and smart home devices.}
}

\newglossaryentry{RSSI}
{
    name={Received Signal Strength Indicator (RSSI)},
    description={Received Signal Strength Indicator (RSSI) is a measure of the relative strength of a radio signal received by a device. Higher values indicate a stronger signal strength. The RSSI is affected by the radio chipset, strength at which the signal is transmitted and environmental factors.}
}

\newglossaryentry{BluetoothBeacon}
{
    name={Bluetooth Beacon},
    description={Bluetooth Beacon is a compact device that can be configured to continuously broadcast an identifier and some custom data as part of a Bluetooth signal. Other Bluetooth-enabled devices can detect these signals to get information, typically specific to the location of the Beacon.}
}

\newglossaryentry{PublicCloud}
{
    name={Public Cloud},
    description={Public Cloud is an Internet-based service that allows users to rent and access remote computation, storage and software capabilities that are hosted at large data centers offered managed by service providers like Microsoft, Amazon, and Google. It reduces the cost and effort in managing physical computing infrastructure at an organization, and at a higher reliability and scalability.}
}

\newglossaryentry{GeoHash}
{
    name={GeoHash},
    description={GeoHash is a mechanism to encode a location in the form of a compact sequence of alphabets and numbers that are easy to remember, compared to latitude and longitude. Typically, longer hashes offer a higher precision of the location.}
}

\newglossaryentry{API}
{
    name={Application Programming Interface (API)},
    description={Application Programming Interface (API) is a description of the input and output parameters that are received and returned when accessing a capability offered by an application.}
}

\newglossaryentry{REST}
{
    name={REST},
    description={Representational State Transfer (REST) is a software architecture that allows desktop and mobile clients to interact with Internet services by passing requests and receiving responses, using web standards such as HTTP and data models like JSON.}
}

\newglossaryentry{TemporalGraph}
{
    name={Temporal Graph},
    description={Like a regular graph, a Temporal Graph (or Temporal Network) is a collection of \emph{vertices} and \emph{edges} between vertices that indicate a relationship between them. But the vertices and edges that exist at different points in time may vary, and their attributes may also change over time. E.g., temporal graphs model interactions in a social network, traffic flow in a road network and proximity contacts in a contact tracing network.}
}

\newglossaryentry{BFS}
{
    name={Breadth First Search (BFS)},
    description={Breadth First Search (BFS) is a graph algorithm designed to visit all vertices that have a path from a particular \emph{source vertex}. It begins by visiting the immediate neighbours of the source vertex, proceeds to visit the immediate neighbours of the newly visited vertices if they are unvisited, and so on.}
}

\newglossaryentry{ReproductionNumber}
{
    name={Reproduction Number},
    description={Reproduction Number $R_0$ of a pandemic is the number of individuals expected to be infected in a population as a direct result of a single person being infected, when all the other individuals are susceptible to the disease, i.e., have not been vaccinated or have not already acquired the disease.}
}

\newglossaryentry{CentralityMeasure}
{
    name={Centrality Measure},
    description={Centrality Measure is a graph-theoretic score that measures the relative importance of vertices in their ability to spread or influence other vertices in the network. Examples of these measures include degree, betweenness, Eigenvalue, closeness centrality, Page Rank, etc. They are used to identify important or critical vertices in contact networks, social networks, WWW graphs, road networks, etc.}
}

\newglossaryentry{QRCode}
{
    name={QR Code},
    description={Quick Response (QR) Code is a 2-D barcode standard which serves as a machine or device readable label that encodes information. Smart phones can use their cameras to take a picture of the QR Code and Apps or libraries can extract the information present in them. Examples of such information include some identifier, the physical location or a URL to a website.}
}

\newglossaryentry{VMs}
{
    name={Virtual Machines (VMs)},
    description={A Virtual Machine (VM) is a computing environment that provides all the functionalities of a full computer, but itself is executing within another computer. A VM is the typical unit of renting a computer in public clouds. VMs help divide a single large computer or server in the cloud into multiple smaller computers, and the VMs are independently rented to different users.}
}

\begin{document}

\title{GoCoronaGo: Privacy Respecting Contact Tracing for COVID-19 Management\thanks{Pre-print of article to appear in the \emph{\href{https://link.springer.com/journal/41745/volumes-and-issues}{Journal of the Indian Institute of Science}}}}
\author{Yogesh Simmhan,
        Tarun Rambha,\\\vspace{-0.3cm}
        Aakash Khochare,
        Shriram Ramesh,
        Animesh Baranawal,\\
        John Varghese George,
        Rahul Atul Bhope,
        Amrita Namtirtha,
        Amritha Sundararajan,
        Sharath Suresh Bhargav\\
        \emph{Indian Institute of Science, Bangalore, India}\\
        \href{mailto:simmhan@iisc.ac.in}{simmhan@iisc.ac.in}, \href{mailto:tarunrambha@iisc.ac.in}{tarunrambha@iisc.ac.in}
        \and
        Nihar Thakkar, Raj Kiran\\\vspace{-0.3cm}
        \emph{Independent Consultants}
}

\date{}

\maketitle


\begin{abstract}
The COVID-19 pandemic is imposing enormous global challenges in managing the spread of the virus. A key pillar to mitigation is contact tracing, which complements testing and isolation. Digital apps for contact tracing using Bluetooth technology available in smartphones have gained prevalence globally. In this article, we discuss various capabilities of such digital contact tracing, and its implication on community safety and individual privacy, among others. We further describe the GoCoronaGo institutional contact tracing app that we have developed, and the conscious and sometimes contrarian design choices we have made. We offer a detailed overview of the app, backend platform and analytics, and our early experiences with deploying the app to over $1000$ users within the Indian Institute of Science campus in Bangalore. We also highlight research opportunities and open challenges for digital contact tracing and analytics over temporal networks constructed from them.
\end{abstract}

\section{Introduction}

Contagious viral diseases such as the SARS-CoV (2002), H1N1 (2009), MERS-CoV (2012), and SARS-CoV-2 (2019) have resulted in global epidemic outbreaks and placed a massive burden on public health systems around the world. These pandemics have cascading effects that result in irreparable consequences to economies and quality of life. The recent SARS-CoV-2 or COVID-19 pandemic has triggered national and regional \emph{lockdowns} across the world to curb the spread of the virus. With incubation periods that last days and with a significant fraction of asymptomatic carriers, the proliferation of the disease has been hard to detect and localize. Further, testing of populations at a large scale has proved challenging due to limited testing kits, well-trained health-care professionals, and funds in emerging economies~\cite{liang2020covid}.


To tackle this problem, governments and health workers use \emph{\Gls{ContactTracing}} of infected individuals to identify those who may have come in contact with them, also called \emph{primary contacts}. These primary contacts are then \emph{quarantined} and/or \emph{tested} depending on their symptoms. Testing, tracing, and isolation form essential components of COVID-19 management, besides \emph{preventive measures} like wearing masks, practising \Gls{SocialDistancing} and washing hands~\cite{Kucharski2020lancet}. 
Traditional methods of contact tracing are often laborious and may be erroneous due to recall biases~\cite{kretzschmar2020impact,10665-332049}. Also, human activity patterns often involve interactions with strangers, especially when travelling, which makes it difficult to identify contacts using traditional methods. 

As a large fraction of the population owns smartphones, countries around the world, including India, have attempted to use \emph{\Gls{DCT}}~\cite{Ferrettieabb6936,10665-332265,Braithwaite-lancet-surv}. Mobile apps that use Bluetooth technology are deployed to record close interactions between users. These Bluetooth Low Energy (BLE) apps typically \emph{advertise} a unique device ID, which can be recognized by other nearby devices with the app that \emph{scan for} and save these advertised IDs, also called \emph{contacts}. This information is typically stored on the local device; if a user tests positive, their Bluetooth contacts are uploaded to a central database and their contacts are alerted. This can dramatically reduce the time required for contact tracing from days to potentially hours, thereby mitigating the spread of the virus~\cite{kretzschmar2020impact}. Examples of such national-scale apps include \emph{Aarogya Setu}\cite{asethu} in India, \emph{TraceTogether}\cite{tracetogether} in Singapore, \emph{COVIDSafe}\cite{covidsafe} in Australia, \emph{COVID Alert}\footnote{COVID Alert App, \url{https://www.canada.ca/en/public-health/services/diseases/coronavirus-disease-covid-19/covid-alert.html}} in Canada, \emph{Corona-Warn-App}\footnote{Corona-Warn-App, \url{https://www.coronawarn.app/en/}} in Germany, etc.

However, there are \emph{limitations} to digital contact tracing. These constraints include the low reliability and asymmetry of Bluetooth technology in detecting nearby users~\cite{Douglas2020,ngpc2020,dyoung2020,dehaye2020}; low accuracy of the proximity distance between users to help distinguish nearby and farther off users~\cite{Douglas2020, ngpc2020}; high degree of adoption required for digital contact tracing to be effective~\cite{mitreview-adoption,hinch2020effective}; and the inability to locate secondary and tertiary contacts until the primary and secondary contacts test positive, respectively. It is hence still important to use complementary digital contact tracing with manual methods.

In this article, we describe \textit{GoCoronaGo (GCG)}~\footnote{Latest details for GoCoronaGo and its download link is available at \url{https://GoCoronaGo.app}}, a digital contact tracing app for institutions, which attempts to address these limitations. A key distinction of our approach is to collect the contact trace data of devices into a centralized database, continuously, irrespective of if or when a person is diagnosed as COVID positive. This proximity data of all app users is used to build a \emph{temporal contact graph}, where vertices are devices, and edges indicate proximity between devices for a certain time period and with a certain Bluetooth signal strength.

This approach has several \emph{benefits}.
When a GCG user is tested positive for COVID-19, we use graph algorithms to rapidly identify primary, secondary, and other higher-order contacts, based on WHO guidelines~\cite{10665-332049}.
Further, even if the Bluetooth scans were missed by the infected user, successful scans by other proximate devices can be used to alert the relevant contacts, increasing the reliability of detection.
In addition, centralized digital contact tracing has the potential to estimate the state of the population using network-based SEIR models, which can be used to assign risk scores and prioritize testing~\cite{stehle11, enright18, koher19}.

Of course, centralized contact data collection has its \emph{downsides}, primarily, the privacy implications of tracking the interactions between a large number of individuals. We take several precautions to mitigate this. One, the app is designed for deployment only within institutions and closed campuses, and not at a city, regional, or national scale. The data collected is owned by the host institution and not by a central authority. Two, users do not have to share any personal information, and devices are identified using a randomly generated ID. Sharing GPS location or their phone number is voluntary and through opt-in. Lastly, deanonymization of data is limited to COVID-19 contact tracing and, by design, requires multiple entities to cooperate, and is overseen by an advisory board with a broad representation from the institution. We discuss these pros and cons in more detail later.

Besides a centralized data collection approach, we also conduct experiments to understand the impact of various smartphone devices and the environment on the Bluetooth signal strength to better ascertain the proximity between devices. We also send proactive messages for users to enable custom Bluetooth settings in their smartphone to improve reliability. The use of the GCG App within an institutional setting, with data collection and usage governed by the organization, may lead to higher adoption of the app, and enhance its effectiveness in contact tracing.

This article examines the design rationale, architecture, and our experience in deploying the GoCoronaGo digital contact tracing app as part of a pilot at the Indian Institute of Science (IISc). It also discusses the challenges and opportunities in improving the utility of digital contact tracing.

The rest of the article is organized as follows: In Section~\ref{sec:background} we review digital contact tracing and provide an overview of a few popular COVID-19 apps. Section~\ref{sec:architecture} provides details of the app design and the backend architecture. In Section~\ref{sec:analytics}, we describe various analytics, including temporal contact network algorithms, for contact tracing, and for providing feedback to app users. 
Finally, Section~\ref{sec:discussion} summarizes our experience with deploying the app at IISc, and highlights some of the opportunities and challenges of digital contact tracing.

\section{Background and Related Work}
\label{sec:background}
\subsection{Contact Tracing}
Infectious diseases, that spread through person-to-person interactions, can be contained by tracking their sources and quarantining the individuals who are or may be affected. This is typically done using physical interviews, which try to determine the places visited and the people met by the patient~\cite{10665-332049}. In some cases, the location history of the patients is shared by cities and public health agencies on websites and mobile apps to allow others who were in the vicinity at that time to take precautions. This form of contact tracing relies heavily on one's memory and collecting such data manually is cumbersome. Contact tracing is crucial, especially for viruses such as the SARS-CoV-2 that exhibit high transmission rates, low testing rates, long incubation times, and a significant fraction of asymptomatic carriers, who could infect other susceptible individuals~\cite{backer20, he20, Ferrettieabb6936}.

Digital contact tracing, on the other hand, involves the use of technology to keep track of the individuals who came in close proximity with each other. It has been shown to be effective in preventing the spread of communicable diseases in livestock~\cite{ortiz06, kao06}, but experiments involving human populations have been limited~\cite{schafer16}. The scale at which COVID-19 has spread has led to the use of Bluetooth and GPS-based contact tracing applications on mobile phones, which help individuals automatically keep a record of the places they visited and the people they met, along with the timestamps. This permits us to build contact neighborhoods that can be used to alert or quarantine the concerned individuals and identify potentially risky interactions. 

\subsection{Digital Contact Tracing for COVID-19}
Most digital contact tracing apps for COVID-19 rely on \Gls{Bluetooth} technology available on smartphones. In addition, a few apps collect the GPS location of users. The rapid spread of the COVID-19 virus has led to the development of a variety of smartphone apps around the world, which are variants on this theme. Examples include both national apps like Aarogya Setu (India), NHSX (UK), and Covid Safe (Australia), as well as those proposed by institutions, like NOVID (CMU) and SafePaths (MIT). A review of contact tracing apps can be found in~\cite{Braithwaite-lancet-surv,ahmed-ieee-surv,li2020decentralized, bassi2020overview}.

At a broad level, these apps scan and advertise for Bluetooth signals and record the timestamp, along with the signal strength or the \Gls{RSSI}, reported in decibel-milliwatts (dBm) in Android~\footnote{ScanResult for Bluetooth LE scan, \url{https://developer.android.com/reference/kotlin/android/bluetooth/le/ScanResult}}. The RSSI values are negative and higher when the devices are close to each other. Translating the Bluetooth RSSI to proximity distances for contact tracing is not straightforward since it depends on numerous factors such as the phone hardware, drivers, operating system, ability to run continuously in the background, and interference due to surfaces. Yet, they have been widely attempted and deployed because of its potential advantages over manual contact tracing. 

In fact, to address some of the interoperability issues across Android phones and iPhones, Google and Apple have even introduced an \emph{exposure notifications (GAEN) protocol} into their OS as part of their COVID-19 response~\cite{google-en}. The BlueTrace protocol~\cite{bay2020bluetrace} used by apps in Singapore and Australia is another popular standard. Europe has two competing contact tracing standards that are being refined, Decentralized Privacy-Preserving Proximity Tracing ($dp3^t$)~\cite{dp3t} and Pan-European Privacy-Preserving Proximity Tracing (PEPP-PT)~\cite{pepp}. 
The Bluetooth Special Interest Group (SIG) is also working on a contact tracing standard for wearables~\cite{btsig}.
Such protocols help with mobility across national boundaries, avoid having to install multiple apps, and in the development of custom, yet interoperable, apps.

Besides smartphone-based apps, others have also developed hardware devices such as the TraceTogether token~\cite{tracetoken} that uses Bluetooth, but operates independently of a phone, or wearables like wristwatches that can track the location using GPS~\cite{bbcwrist}.
In addition to Bluetooth, a few apps like NOVID also broadcast ultrasound signals using a phone's speakers and other apps in the vicinity detect them using their microphone~\cite{novid}.
There have also been other digital apps such as the NZ COVID Tracer that use QR codes for users to check-in when they enter specific locations\cite{nz-covid}.
Besides contact tracing, digital tools have also been used to track symptoms among populations to identify emerging ``hotspots'' and for health professionals and volunteers to coordinate their response~\cite{10665-332265}.

\begin{figure}
\centering
	\begin{sideways}
\begin{minipage}{\textheight}
	\captionof{table}{Table comparing GCG features with other COVID Contact Tracing Apps, as on \emph{9 Sep 2020}}
	\label{fig:compare}
	\includegraphics[width=1\textwidth]{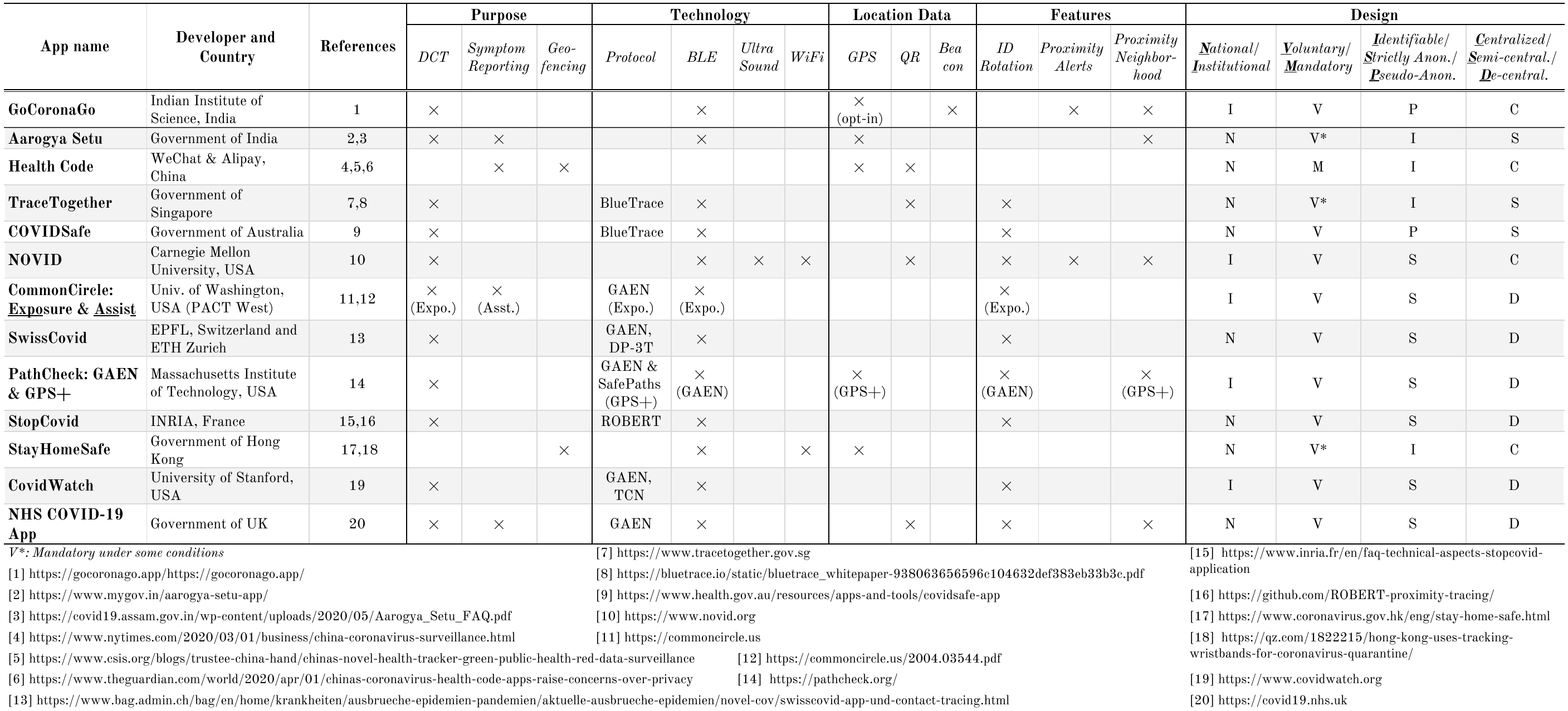}
	\end{minipage}
\end{sideways}
\end{figure}

However, the global adoption of contact tracing apps is low. The percentage of the population who have installed such apps has struggled to go past 20\%, even among developed countries where a majority of the individuals have smartphones~\cite{sensortower}. While there is debate on the minimal adoption rate required for contact tracing apps to have a tangible effect, some use is better than none and more is better~\cite{mitreview-adoption,hinch2020effective, kretzschmar2020impact}. In particular, higher adoption rates in dense neighbourhoods can highlight the effectiveness of tracing effective since the risk of spreading the infection is greater in closely-knit communities. 

\subsection{Balancing Community Safety and Individual Privacy}
There are a number of ways in which one can design such digital contact tracing apps. These offer different trade-offs in terms of individual privacy and the health and safety of the community.

\subsubsection{National vs. Institutional Use}
The target of the app may be for \emph{national/regional use or institutional use}. While national-scale contact tracing apps potentially offer greater ability to manage the pandemic, they also carry greater risks of data leaks and misuse~\cite{china-nyt}. Further, a high degree of adoption at such large scales is challenging, limiting the usefulness of the app for contact tracing. Apps deployed at an institutional scale can be better targeted to the audience and offer better uptake due to the fact that the data is managed at the organizational level. Institutions can also respond more rapidly based on insights provided by the app. But they are less effective when users are moving outside the confines of campuses and interacting with the broader population. E.g., apps like Aarogya Setu and TraceTogether are national apps, while GoCoronaGo, NOVID, and Covid Watch are designed for institutions.

\subsubsection{Voluntary vs. Mandatory}
The use of the app may be \emph{voluntary or mandatory}. Some countries like China have made such apps mandatory for all residents, or for those meeting certain requirements such as travelers. Even organizations may make such national or institutional apps mandatory within their premises. But most countries and institutions tend to keep the use of such apps voluntary. Further, the use of the collected data for contact tracing may also be voluntary or mandatory. If voluntary, there is an explicit opt-in by the individual who is tested COVID positive or is quarantined, before contact tracing using their data can be initiated. Alternatively, there may be rules in place that allow the government or institutions to use any proximity data that is available with them, without additional consent from infected users. An explicit consent helps address concerns of social stigma around COVID patients.
The use of GCG is strictly voluntary, and there is an additional consent required by a user who is infected with COVID-19 before their data can be used for contact tracing -- this, despite their data already being available centrally in the backend.

\subsubsection{Identifiable vs. Anonymized}
Apps may collect \emph{identifiable, strictly-anonymous, or pseudo-anonymous information} as part of contact tracing. 
Some apps like Singapore's TraceTogether compulsorily require the contact details and/or a national identification number to be shared when installing the app. This makes it quicker to reach-out to users during contact tracing, but also heightens the risk of misusing the data for the surveillance of specific individuals and can lead to a significant loss of privacy if the data is breached. In a strictly anonymous setting, no personal information of the user is collected, and they are only identified by a random ID, which itself may also be changed (or ``rotated'') periodically. A set of such IDs may be provided by a central server (TraceTogether) or generated locally by the App. During contact tracing, the user's app is alerted and they have the option of voluntarily responding by contacting the health center or a government agency. If the user uninstalls the app, it may be impossible to do contact tracing. A hybrid approach of pseudo-anonymization ensures that the contact trace data itself is anonymous, but the information required for de-anonymization is available with a trusted independent authority whose consent is required (optionally, with a consent from the infected individual) to identify the users relevant for contact tracing. GCG adopts this hybrid model that balances the privacy of users while also enabling rapid and reliable outreach during contact tracing.

\subsubsection{Centralized vs. De-centralized}
The contact tracing data may be kept \emph{de-centralized, semi-centralized, or centralized}. If de-centralized, the Bluetooth device IDs observed by a user's app are stored locally on the device. 
When a user tests positive for COVID-19 they can inform a backend service of their device ID (potentially, multiple IDs, in case of ID rotation) and their status. The backend periodically relays a list of device IDs associated with COVID positive individuals to all apps, which is then used by the user to verify if they came in contact with a COVID positive person. This is used by PACT~\cite{rivest2020pact} and Google-Apple exposure notification (GAEN) framework~\cite{megan-spectrum-2020}. 

In a semi-centralized approach, a mapping between an app and its device ID is maintained centrally, but the contact trace data remains locally on the device. On testing positive, a user may choose to (or be required to) upload the contact trace data for the recent past to a backend service, which then sends notifications to these primary contact devices asking them to quarantine or get tested.  Examples of this approach include BlueTrace~\cite{bay2020bluetrace} and Aarogya Setu~\cite{asethu}. However, Aarogya Setu also allows users to voluntarily upload their Bluetooth contact data to central servers at any time to get an estimate of other high-risk users in the vicinity.

Lastly, in a centralized approach, both the mapping of apps to device IDs as well as their contacts are sent to a backend service periodically. When a user reports themselves as COVID positive, contact tracing can be initiated on the centralized data already available, optionally after an additional consent. GCG adopts this model.
This variant is relatively intrusive, but arguably has advantages that may justify its use.
One, contact data from both the infected and the proximate users can be combined to increase the reliability of contact tracing. Two, even if users uninstall the app, if the data collected is personalized or is de-anonymizable, then contact tracing can still happen over the backend data for the period during which the app was kept installed. Three, not just primary but even secondary and tertiary contact tracing can be performed rapidly. And four, having a centralized model allows us to perform temporal analytics on a global contact network. This can help identify high-risk individuals for prioritizing preventive, testing and (future) vaccination strategies, and infer the health of the user population.

\subsubsection{Location data and longevity}
Bluetooth data provides the relative interaction between proximate users but in itself does not reveal the spatial location of users. While this may disclose interaction patterns between (anonymous) users, which is necessary for contact tracing, correlating this with particular individuals is not possible without additional out-of-band knowledge about them.

Some contact tracing apps may also collect GPS data (COVID SafePaths) and data from beacons or QR codes (NOVID) that may reveal the absolute spatial location of the users. Collecting spatial location has some benefits. The coronavirus may be transmitted through surfaces or be suspended in the air, and thereby be passed on to others who are not near an infected user but in the same location soon after~\cite{van2020aerosol}. Bluetooth based proximity will miss such users. Also, GPS data collection may be more reliable than Bluetooth.
However, GPS is not precise enough to be useful for identifying proximity between users. Furthermore, tracking the spatial movements of users continuously can have serious privacy consequences~\cite{uber-cnn,hrw-location}. \Gls{BluetoothBeacon}s and scanning \Gls{QRCode}s present at well-known locations can also provide such spatial information, but will be limited to places where the beacons or codes are deployed. GCG allows users to optionally share their GPS data through an explicit opt-in, and also allows the selective use of beacons deployed by institutions.

Lastly, we need to consider the duration for which the centralized or de-centralized data that is collected is retained. This needs to be explicitly stated by the apps for transparency. More the data that is collected and more personalized it is, greater are the consequences for retaining it longer, especially in a centralized or semi-centralized setting. Typically, the contact trace data itself is useful only for roughly 30 days after it is collected since this duration is typically the outer time-window of transmission of the virus. Also, there should be clarity on how long the data is retained after a user uninstalls the app.
GCG deletes a user's phone number, the only personal data they may share, from its backend within 3 months of them uninstalling the app. The anonymized contact trace data is retained for future research purposes, as per the rules set out by the Institute Human Ethics Committee (IHEC).

\section{GCG Architecture}
\label{sec:architecture}
\begin{figure}[t]
	\centering
	\includegraphics[width=1\textwidth]{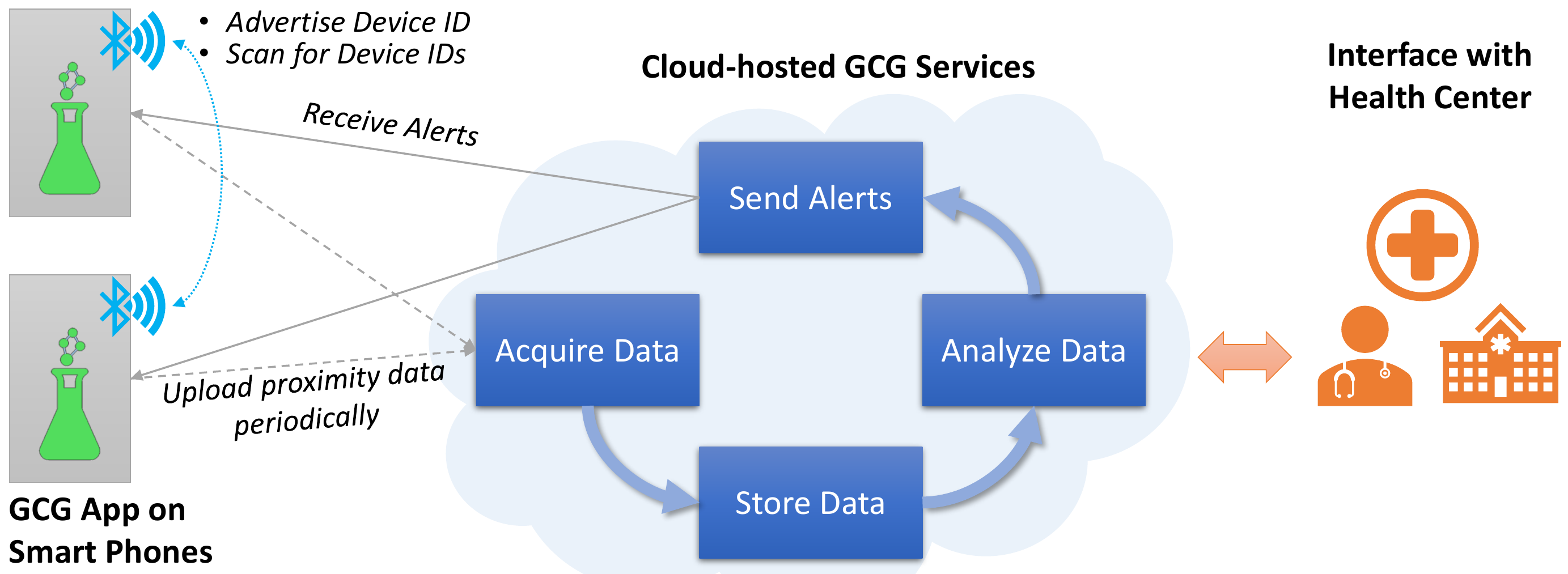}
	\caption{Overall Design of GCG}
	\label{fig:design}
\end{figure}

The GoCoronaGo (GCG) contact tracing platform consists of a smartphone app and backend services for data collection, management, and analysis. The app is designed for COVID-19 \emph{operations and management} within an institution, and is also proposed as a \emph{research project} governed by the Institute Human Ethics Committee (IHEC). The design and technical details of the app and the backend services are described in this section. A high-level design is illustrated in Figure~\ref{fig:design}.

\subsection{Design of the GCG Smart Phone App}
\subsubsection{App Installation and User Registration}
\begin{figure}[t]
	\centering
	\includegraphics[width=0.9\textwidth]{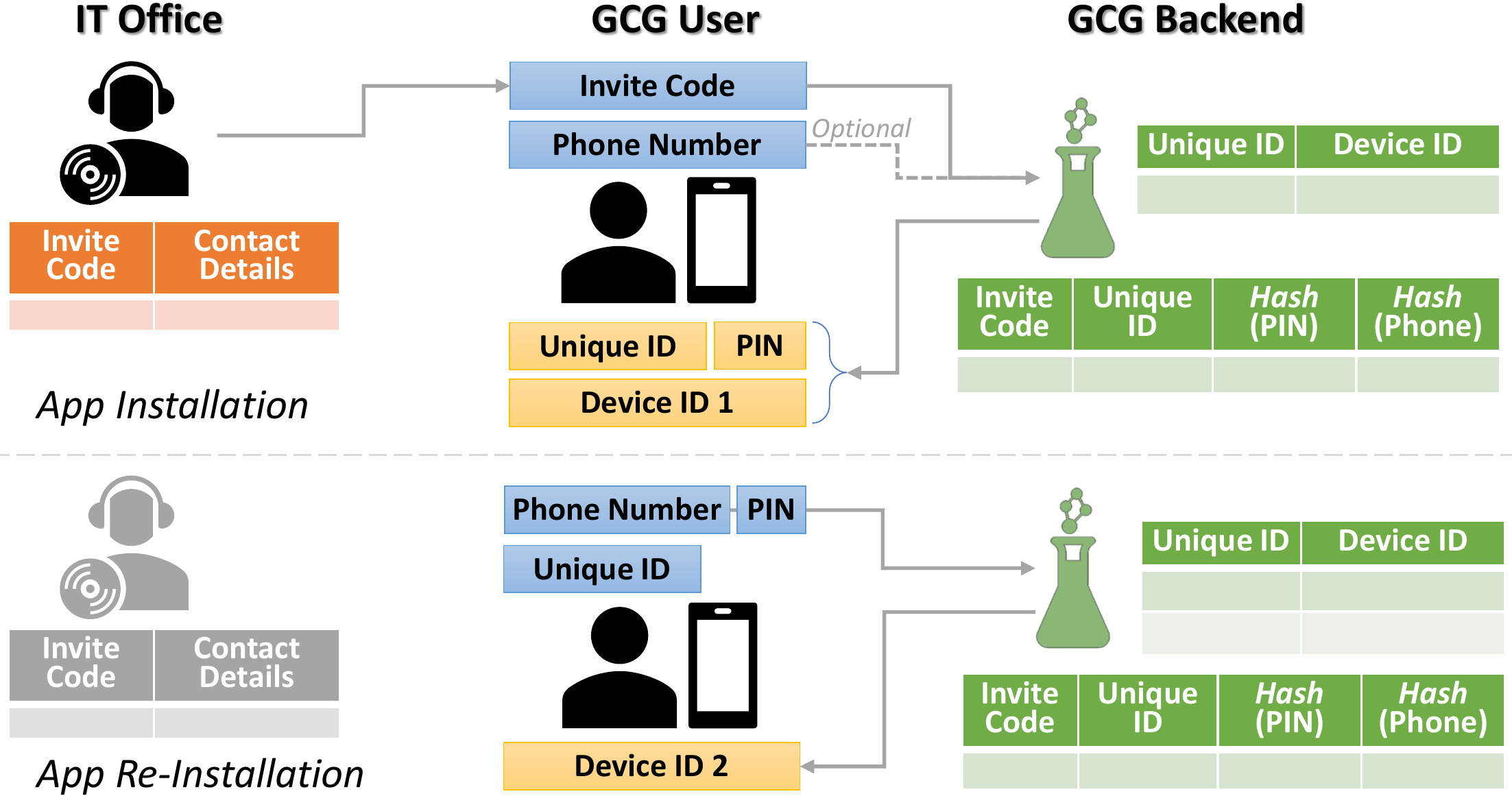}
	\caption{Identifier mapping during GCG App Installation}
	\label{fig:install}
\end{figure}

The GCG App is limited for use by authorized institutions. Since not all institutions may have a private/enterprise app store for their organizations, hosting the app in the public Google Play or Apple App store is convenient. Users at authorized institutions are provided with individual \emph{invitation codes} by a separate entity within the institution, typically the information technology (IT) office. The IT office also maintains a \emph{mapping} from the user's unique \emph{invite code} to the actual individual to whom the code was provided, along with their \emph{contact details}, as shown in Figure~\ref{fig:install}. This mapping from the individual to their invitation code is later used by the IT office during contact tracing, as described in Section~\ref{sec:tracing}.
The user can download the GCG App from the Google Play Store or from an institutional download link. During installation, users enter this invite code into the app, which submits and validates it with the GCG backend servers and is returned a \emph{unique ID}, a \emph{device ID}, and a \emph{PIN}.

The GCG backend maintains the \emph{mapping} from the invite code to the unique ID for the installed device. The invitation code can only be used once by the user for the first installation. To allow future re-installations, a \emph{PIN} is generated for this invitation code and is shared with the user.
Optionally, the user may provide their one-time password (OTP)-verified \emph{phone number} during installation, which is recorded in the backend. This number can be used along with the PIN to reinstall the app in the future, in place of the one-time-use invite code. Lastly, a \emph{device ID} in the form of a random $128~bit$ UUID is generated by the backend for each re/installation on a phone, and a \emph{mapping} is maintained from the unique ID to the device ID, along with the creation timestamp. This device ID will be broadcast as part of the Bluetooth advertisement (Figure~\ref{fig:install}). Both the invite code to unique ID, and unique ID to device ID mappings are used during contact tracing (Section~\ref{sec:tracing}).

A final piece of information collected from the app during re/installation is the \emph{make and model} of the phone. As we discuss later, this is vital for interpreting the Bluetooth signal strength and translating it into a distance estimate.

These identifiers are designed to maintain the anonymity of users from the GCG App and backend, enable de-anonymization of contact users upon an authorized request for contact tracing, and ensure that the app can be re/installed by authorized users. Such sandboxing and identifier-indirection ensures that no single entity -- the IT Office, a GCG user, or the GCG backend -- can independently find the identity of any (other) user and their trace.

A key tenet of GCG is transparency. The installation process in the GCG App has disclosures on the \emph{legal terms and conditions} for the use of the app, and on how the data collected will be used. In addition, there is also a multi-lingual \emph{informed consent}, as required by IHEC, which clearly documents the scope of the research project, potential benefits and downsides, voluntary participation, etc. 

\subsubsection{BLE Advertisement and Scanning}
The GCG App uses \emph{\Gls{BLE}} signals to detect other proximate phones running the app. The \emph{BLE} wireless protocol is ubiquitous among smartphones sold within the last 6 years. It enables low power, short-range wireless communication, and is intended for applications in fitness, smart homes, healthcare, beacons, etc. Its maximum range is $<100~m$~\cite{Bertuletti2016} though this is affected by environmental conditions and transmitting power, and $\approx 10~m$ is the typical range~\cite{Douglas2020}.

BLE devices use an \emph{advertising and scanning protocol} to discover each other and establish a connection. When acting as a \emph{server}, the devices advertise one or more \emph{services} that they support, which are identified by \emph{service assigned numbers};  when acting as a \emph{client}, they find servers to connect, to based on the advertised service assigned numbers.\footnote{Bluetooth GATT Service Assigned Numbers, \url{https://www.bluetooth.com/specifications/gatt/services/}} A single device may advertise multiple services, and it can include a custom payload such as a service name. Also, the BLE advertisement is broadcast in an open channel, which any nearby BLE client can discover. Besides standard $16~bit$ service numbers that are registered and pre-defined for specific types of services, applications can also generate and use $128~bit$ UUIDs for custom services they provide. Once discovered, clients can establish a network connection with the service to perform additional operations such as data exchange.

The GCG App acts as both a client and a server when using the scanning and advertising capabilities of BLE, respectively.  Specifically, it advertises two service assigned numbers, \texttt{0x1800} which represents a \emph{Generic Access} service, and another custom service whose assigned number is the unique \emph{device ID} for a particular app installation. This advertisement is broadcast continuously. As a client, the GCG App scans for $5~secs$ in every minute for advertisements that contain the service number \texttt{0x1800}. If found, it extracts and records the device ID that is sent as a secondary service number in the same advertisement. Piggy-backing the device ID as a service assigned number rather than a custom payload takes fewer bytes, which in turn can reduce the power consumption for the advertisement.

As part of the scanning, the GCG App also retrieves the \emph{Received Signal Strength Indicator (RSSI)}, which is the strength of the BLE signal that is received by the app. As we discuss later, this can be used to estimate the proximity distance.

The GCG Android App uses the default BLE settings for broadcasting its advertisements\footnote{AdvertisingSetParameters, Google Developers, \url{https://developer.android.com/reference/android/bluetooth/le/AdvertisingSetParameters}}, which translates to BLE broadcasts every $1~sec$ at a medium transmission power level.
Also, the app consciously does not establish a connection with apps on another device; the device ID is broadcast to any BLE device that is in the vicinity. In fact, we explicitly set the \emph{connectable} flag on the advertisement to \emph{false}. This enhances security by avoiding malicious content from being transferred.

\subsubsection{Support for GPS and beacon locations}

While such proximity tracking is helpful for contact tracing of individuals who were spatiotemporally co-located, this does not address situations where two users shared the same space, such as an ATM, mess dining hall, or campus grocery, but for a short time apart. Since COVID-19 can be transmitted through surfaces and can linger in the air for some time~\cite{van2020aerosol}, it is beneficial to identify users who were in the same location but not at the same time, especially for locations with a lot of footfall. 

The GCG App allows users to voluntarily share their \emph{GPS location} information with the backend. This is disabled by default. If enabled by the user, the GPS location is retrieved and uploaded to the backend every $5~mins$, and buffered for retries.

Since the sharing of GPS location is strictly voluntary, GCG supports the selective use of \emph{beacons} installed by institutions at such high-risk spaces. These beacons behave like a GCG App that passively advertises its device ID, and the smartphone app can scan for and record the beacon's ID, just as it would detect another GCG smart phone's device ID. Specifically, we use the \emph{iBeacon} protocol from Apple. The beacon transmits a static GCG UUID as its service number, \texttt{0x004C}, as the manufacturer ID for the protocol, and a \emph{major} and \emph{minor version number} to uniquely identify that beacon.
The GCG App scans for the static service number, filters results based on the manufacturer ID, and retrieves the major and minor version numbers. The app encodes these version numbers into a template UUID to form a unique device ID for that beacon and adds it to its proximity trace.

\subsubsection{Buffering proximity data for reliability}
During each scan, the proximity data collected consists of zero or more device ID(s) and the corresponding RSSI values that were discovered at that timestamp. Performing a service call to send this data to the backend servers consumes power and bandwidth on the phone. Instead of sending this data after each scan, we buffer it to a \emph{SQlite} database on the phone and periodically send the buffered data to the backend in a single batch. This transmission interval is set to $15~mins$. This type of batching amortizes the power and network costs across scans, while ensuring the freshness of the data available at the backend. 
Buffering is also beneficial when Internet connectivity is intermittent. If the proximity data cannot be sent to the backend, the buffered data is retained on the device and a resend attempt is made in the next transmission interval. 

Given that this is the most frequent service call to the backend, we use a compact \emph{binary serialization} to represent the proximity data sent to the backend, unlike the other services which use JSON. \ysnoted{Typically, we observe that each device sends a median of \Note{20 proximity entries and ??? bytes of cumulative data} per day.\ysnote{AK/RB to fill in}\drnote{AK: RB do you have an estimate? If not ping me. I'll look into it.}}

\subsubsection{Telemetry for app health monitoring}
The GCG App needs to run in the background all the time for effective Bluetooth advertising, scanning, and proximity data collection. However, the heterogeneity of smartphone models and the limitations of their OS means that this advertising and scanning may not be reliable. To identify issues with specific device models and app installations, and verify if the app is running, we collect and report \emph{liveliness} telemetry statistics to the backend every hour.
These include a count of BLE scans performed, BLE scans failed, GPS scans, GCG users and beacons detected, and contact buffer size; Bluetooth and GPS enabled status, Bluetooth and GPS permission flags, battery level, app version, etc.
These statistics also help us in understanding the aggregate usage of the GCG App within an institution.

\begin{figure}%
    \centering
    \subfloat[Main screen]{\label{fig:ui:main}
        \includegraphics[width=0.31\textwidth]{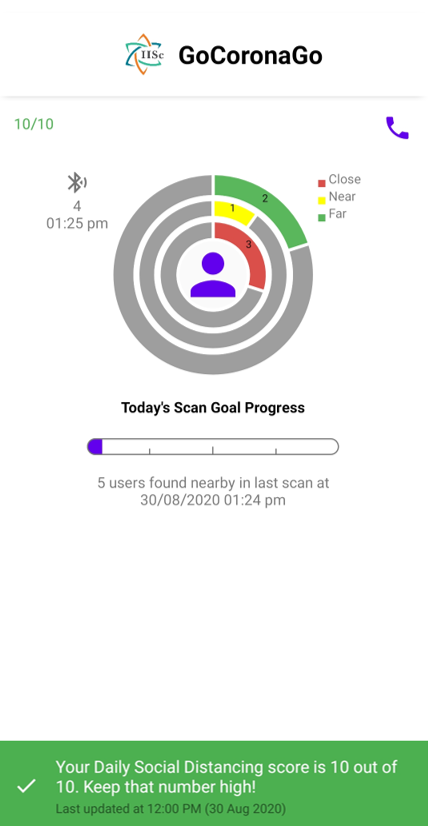}    
        \vspace{0.2cm}
    }~
    \subfloat[Hourly contacts]{\label{fig:ui:contacts}
        \includegraphics[width=0.295\textwidth]{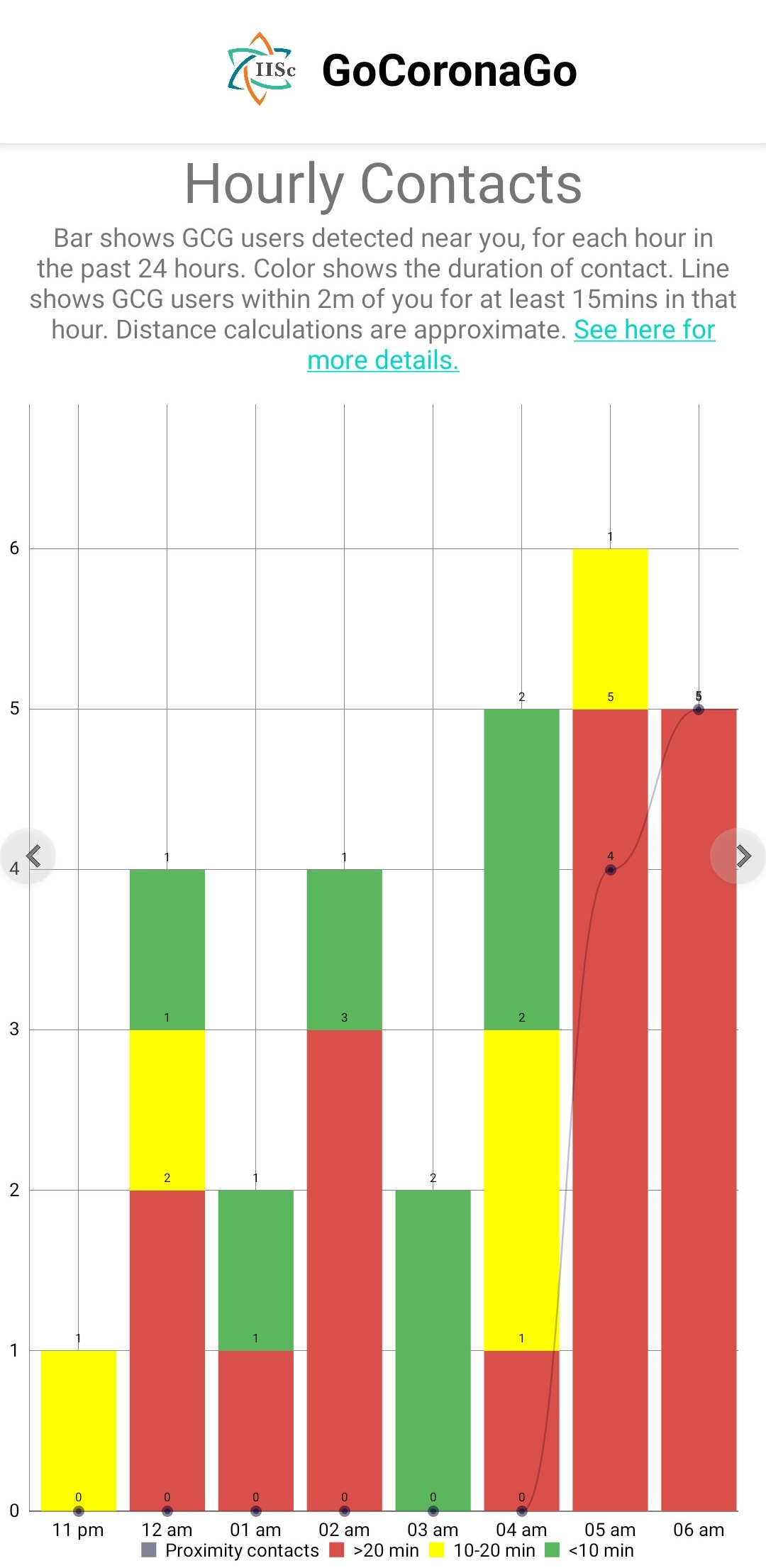}    
    }~
    \subfloat[Hourly scans]{\label{fig:ui:scans}
        \includegraphics[width=0.295\textwidth]{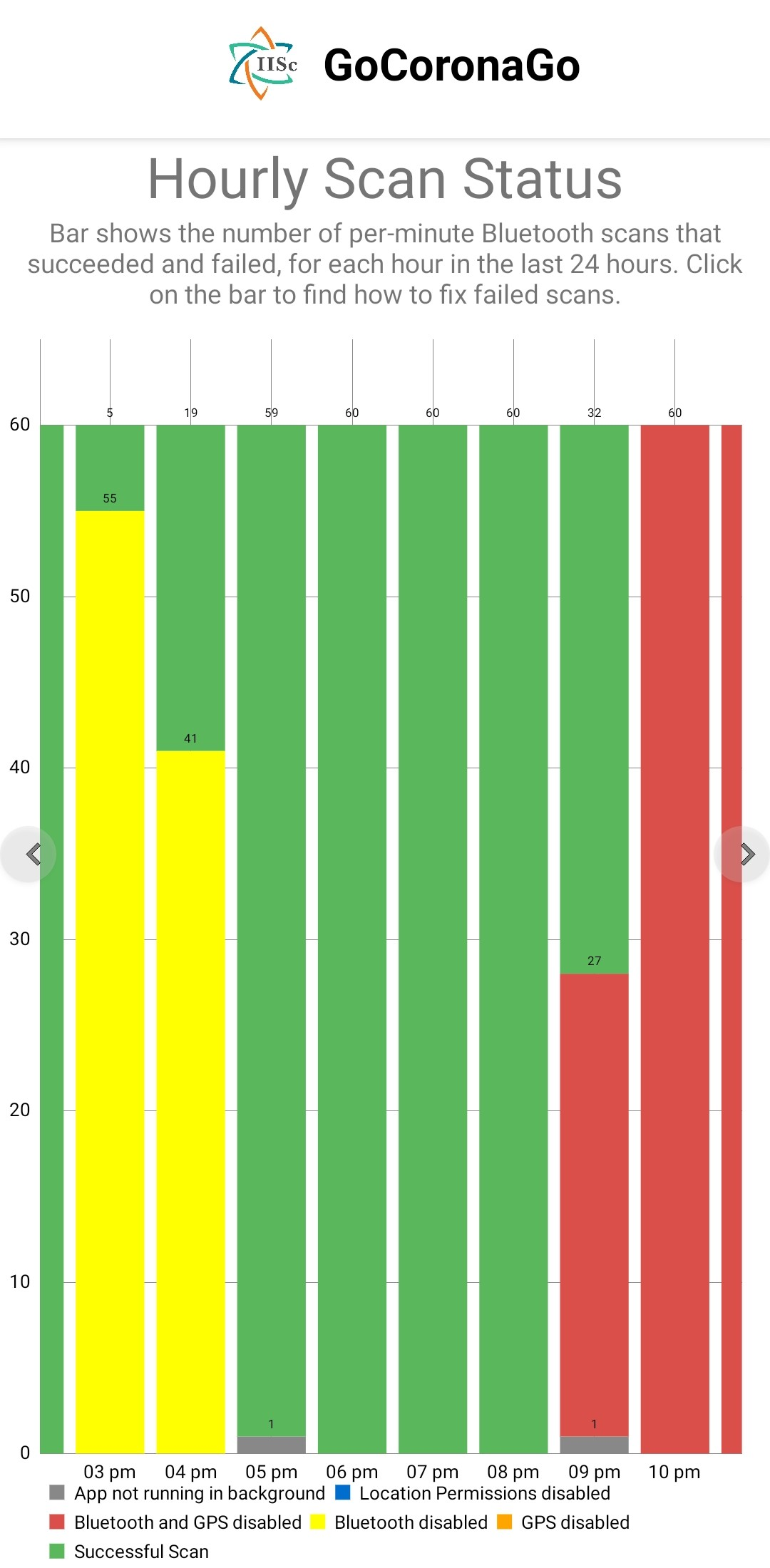}    
    }\\
    \subfloat[Device Density Heatmap]{\label{fig:ui:density}
        \includegraphics[width=0.3\textwidth]{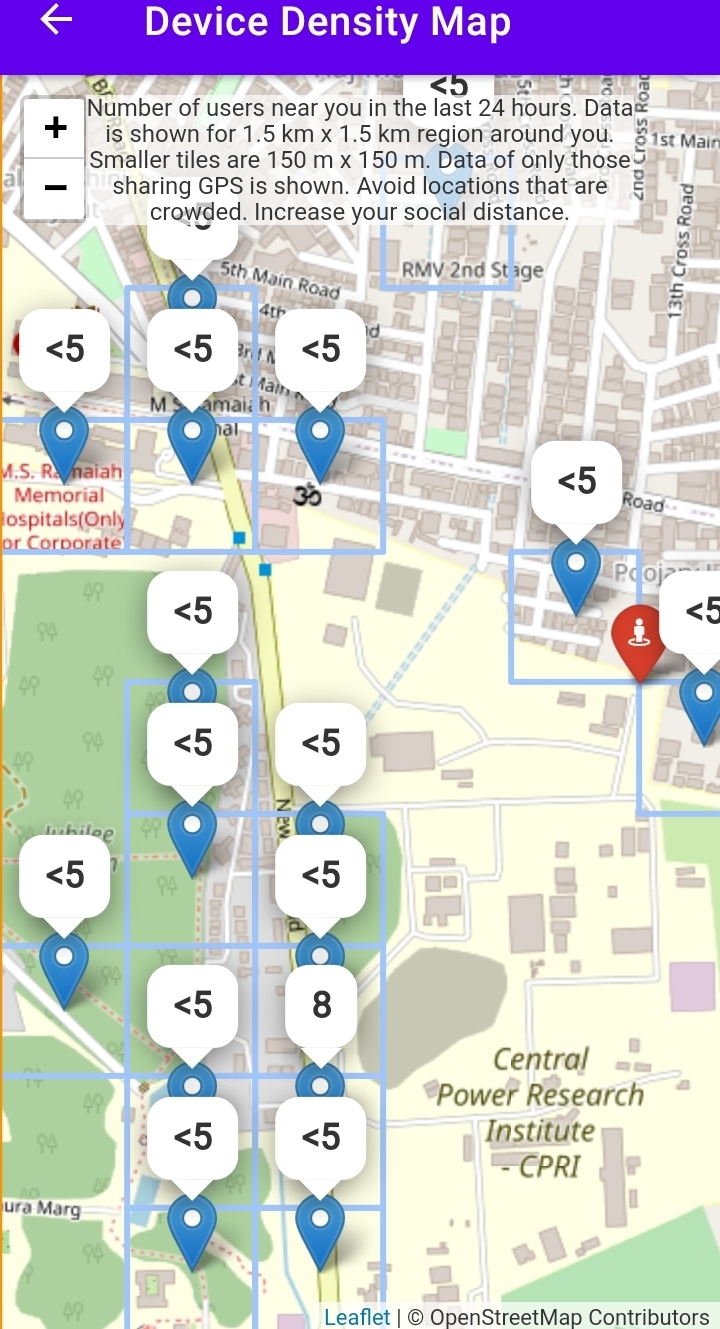}    
    }~
    \subfloat[Contact Network]{\label{fig:ui:nw}
        \includegraphics[width=0.3\textwidth]{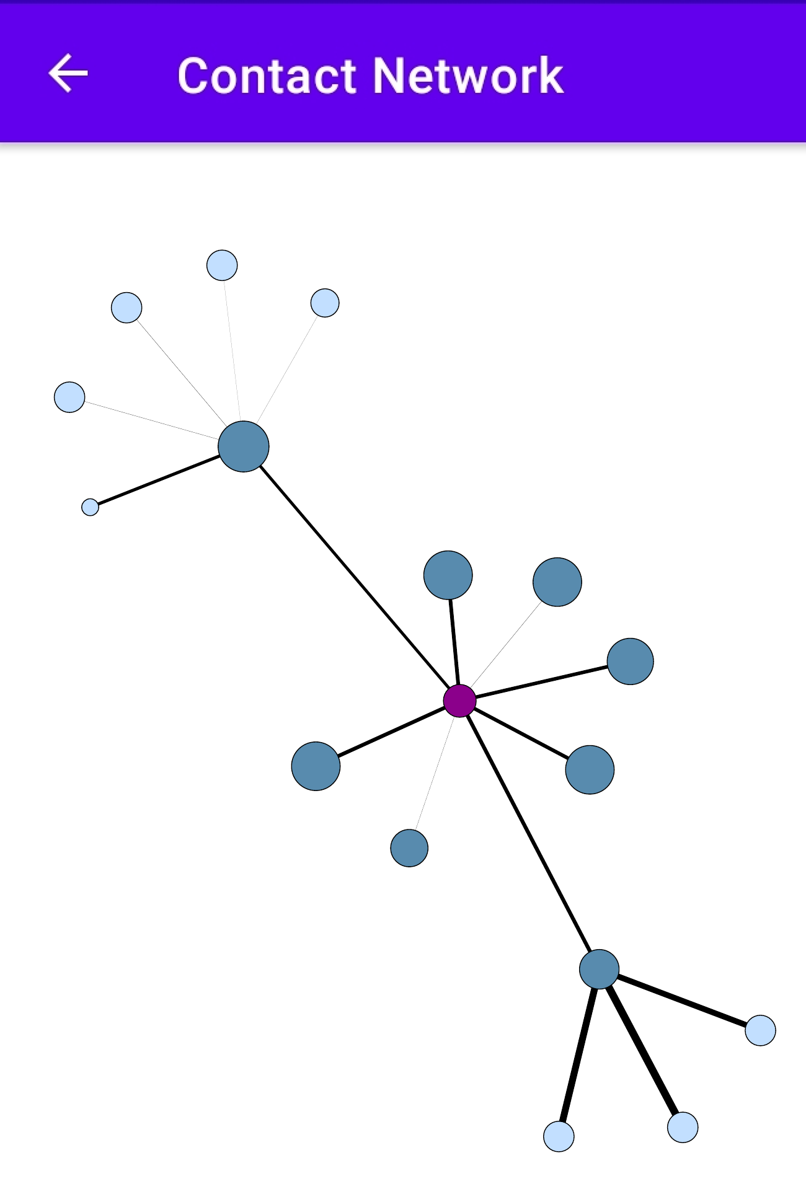} \vspace{1.3cm}
    }~
    \subfloat[Alerts panel]{\label{fig:ui:alerts}
        \includegraphics[width=0.3\textwidth]{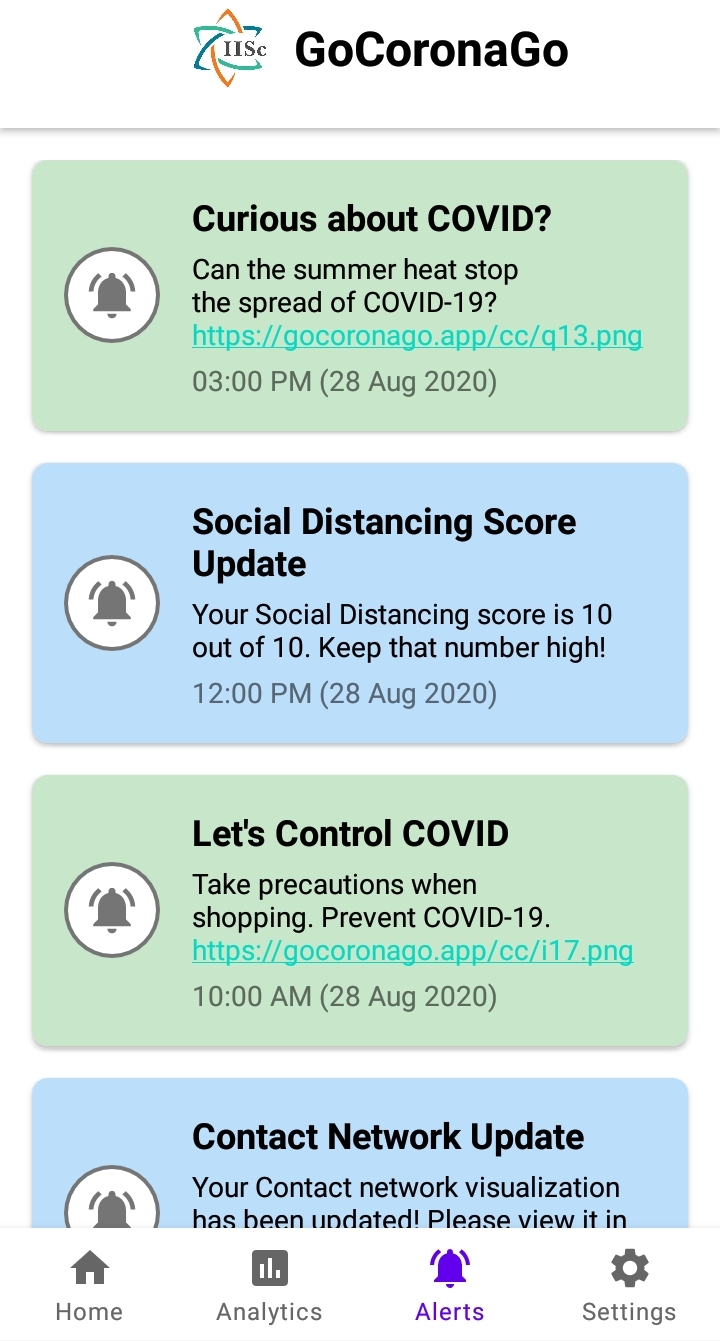}    
    }
    \caption{User interface and analytics in the GoCoronaGo v0.7 Android App}%
	\label{fig:ui}
\end{figure}
\subsubsection{UI and Analytics}
Besides tracking Bluetooth contact data, the GCG App offers several features to inform the users about COVID-19 and engage them in preventing its spread. Screenshots of these UI elements are shown in Figure~\ref{fig:ui}.

Key among these is a \emph{Proximity Alert}, wherein a notification is triggered on the smartphone if $5$ or more users (configurable) were detected within a $\approx 2~m$ distance during the last Bluetooth scan. This acts as a warning to users in case they inadvertently overlook social distancing. As discussed later, the $2~m$ distance threshold is just an estimate based on the RSSI. The alert is also triggered only once an hour (configurable) to avoid saturating the user.

In addition, users can visualize a plot of the \emph{hourly count of contacts} segregated by the duration of contact within the hour, e.g., $<10~mins$, $10-20~mins$ and $>20~mins$ (Figure~\ref{fig:ui:contacts}). This gives them a sense of their interaction pattern for the past $24~hours$. Similarly, we also display the number of \emph{scans performed each hour} for the past $24~hours$ (Figure~\ref{fig:ui:scans}). This can help identify issues with Bluetooth scanning on specific phones, and prompts the user to take corrective measures. A summary of the number of scans completed per day is also shown as a \emph{progress bar} to motivate users to hit $1000$ or more of the $1440$ possible $1~min$ scans (Figure~\ref{fig:ui:main}).

These \emph{local analytics} within the app are complemented by \emph{aggregate analytics} performed in the backend and are shared through the app each day. These include the \emph{social distancing score}, \emph{user density heatmap} for neighboring locations, and a visualization of the \emph{contact network neighborhood}. These are described later in Section \ref{sec:analytics}.
A unique aspect of the app is that the set of remote analytics available can be dynamically changed without having to update the app. In the future, this can also be used to push forms and conduct surveys from within the app.

Importantly, none of the analytics provided to users reveals the identity of other users or even their device IDs, to \textit{protect their privacy}. E.g., the hourly contact bars only report the aggregate counts of nearby devices and cumulative duration of interaction at different distances, while the proximity alert is triggered only if at least 3 or more users are nearby to prevent fine-grained estimates of even the number of GCG users.

Lastly, we also provide helpful information to educate users about COVID-19. These include a \emph{plot} of the positive, recovered, and deceased cases across time in India, and in the local state, and a \emph{map} of the current positive cases at the state and district level. In addition, we also share \emph{Let's Control COVID} and \emph{Curious about COVID?} infographics as app alerts each day, which suggest precautions, debunk myths, and offer scientific information (Figure~\ref{fig:ui:alerts}). These are sourced from public health and science resources such as WHO~\footnote{Information for the public, World Health Organization, \url{https://www.who.int/westernpacific/emergencies/covid-19/information}}, the COVID Gyan initiative from IISc-TIFR~\footnote{COVID Gyan, TIFR and IISc, \url{https://covid-gyan.in/}}, and Indian Scientists' Response to COVID-19~\footnote{Indian Scientists' Response to CoViD-19, \url{https://indscicov.in/}}.

\subsubsection{Android and iOS implementations}
The features described here are largely applicable to \emph{GoCoronaGo v0.7} on \emph{Android} smartphones. \emph{GoCoronaGo v0.2} is a lighter version available for \emph{iOS} with features limited to advertising, scanning, and receiving alerts. This is due to the limited numbers of iPhone users on the academic campus.

There are other OS and device-specific issues as well that we encountered and addressed in various iterations of the app. While we were initially using wildcard filters when performing Bluetooth scans for service numbers on the Android app, we noticed that certain phone models such as Samsung did not reliably perform such scans. This led us to adopt the \texttt{0x1800} approach. 

Continuous Bluetooth advertisement and scanning in the background is challenging in Android, and virtually impossible in iOS. 
Smartphone brands with custom Android builds, such as Xiaomi, Oppo, Vivo, etc. do not always support the recommended practise of executing such applications as a \emph{foreground service} with a persistent, ongoing notification~\footnote{Services overview, Google Developers \url{https://developer.android.com/guide/components/services}}. As a result, users are forced to change the Android battery usage settings and/or autostart permissions for the GCG App, which are brand and even model specific. Absence of reliable scanning and advertising defeats the key purpose of the app. We provide local analytics and alerts to help users address such issues.
Further, Android requires users to enable GPS to even perform continuous Bluetooth scanning, as a way to indicate to users that their location may be revealed indirectly, say, through beacons at well-known locations. But requiring GPS to be on even though the app does not collect the GPS location without opt-in confuses users, and may lead to privacy concerns.

On iOS, the problems with background Bluetooth advertisement and scanning is well documented due to Apple's restrictive policies~\cite{wired-nhsx,smh-aus,bay2020bluetrace}. The iOS GCG App is effective when in the foreground and the user is viewing the app. However, when the user is not actively using the app or the phone is locked, the app can scan for other devices that are advertising, but it cannot advertise. As a result, there needs to be other Android or active iOS GCG devices nearby for contacts to be recorded, colloquially referred to as ``Android Herd Immunity''~\cite{guard-nhsx}.

Besides technical challenges, there are also policy challenges in deploying COVID-19 related Android and iOS apps to the Google Play and Apple App stores. Certification from an official Government of India agency with specific verbiage was required before the GCG Android app would even be reviewed for hosting on the Play store, and the subsequent reviews of the app's updates take a week or longer. Given the restrictions that Apple imposes on apps posted on its App Store, the iOS GCG App is only viable for an \emph{ad hoc} or enterprise license deployment.

\subsection{Design of the GCG Backend Services}

\begin{figure}[t]
	\centering
	\includegraphics[width=0.9\textwidth]{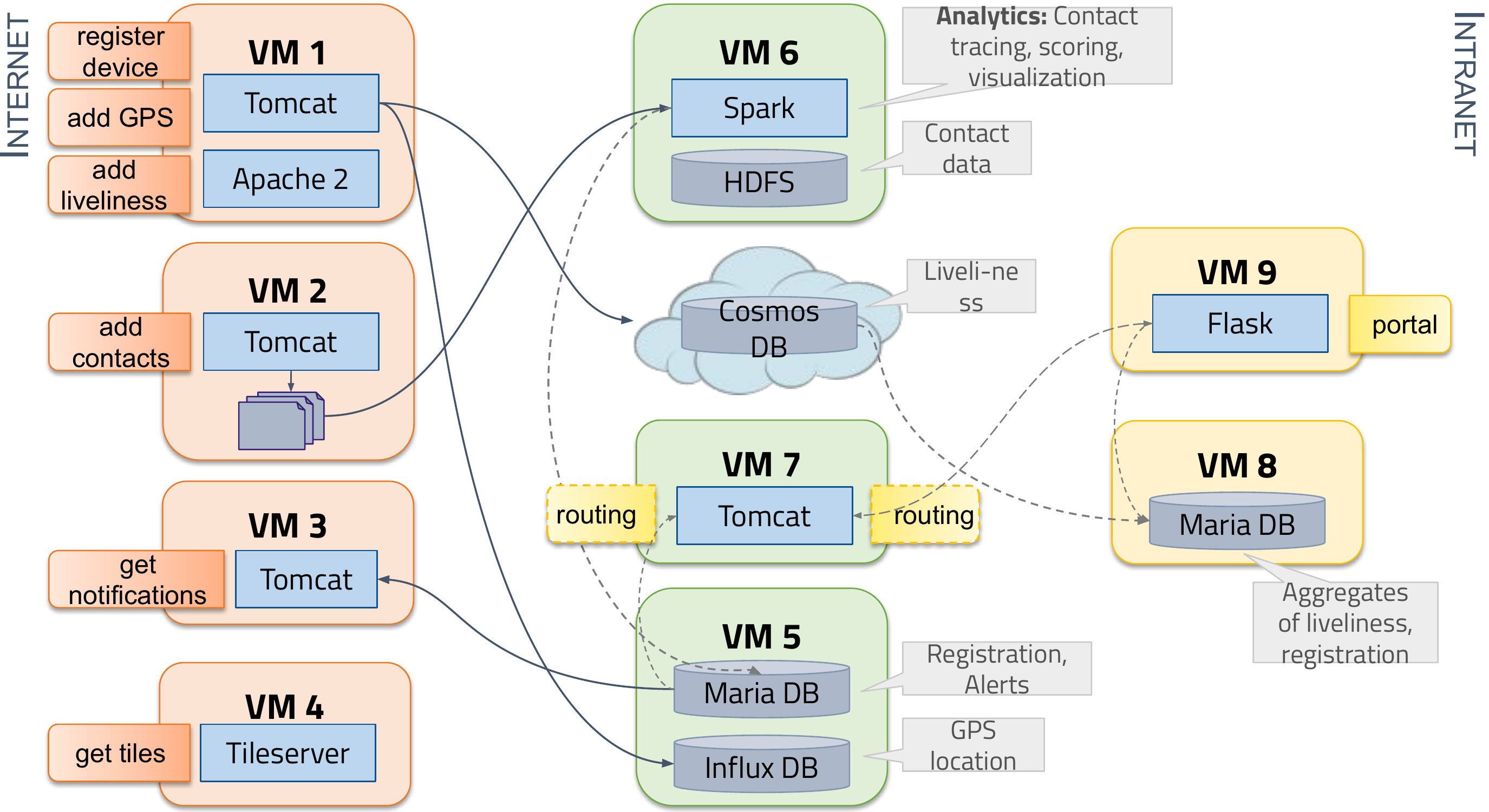}
	\caption{Backend VMs, services and databases, and their interactions}
	\label{fig:arch}
\end{figure}

GCG web services, data management, and analytics are hosted on the Microsoft Azure \Gls{PublicCloud}. As shown in Figure~\ref{fig:arch}, these are present on different \Gls{VMs} that are segregated based on their \emph{workload} (service endpoint, data management, analytics), and their \emph{security zone} (Internet, Intranet, and internal). We describe these backend capabilities next.

\subsubsection{Internet-facing services}

A suite of \Gls{REST} service \Gls{API} is defined for the GCG App to interface with the backend, to upload data and to download analytics and alerts. The REST services are implemented using Java Servlets running on Apache Tomcat Web Server, and their service endpoints are accessible on the Internet. These APIs include \emph{register device, add proximity contacts, add GPS, add liveliness, get notifications}, and \emph{fetch analytics}. Most use JSON as the REST body, except \emph{add contacts} which uses a binary protocol.

The \emph{register device} API accepts an invitation code from the app, checks a MariaDB table if the code is present, not expired and not yet used, and if so, generates a random device UUID, a random PIN and a unique ID for the user, which are returned back to the app. These mappings, as described earlier, are maintained in MariaDB. The phone number, if provided, is salted, hashed and stored in the database for comparison in the future if a user reinstalls the app. The number is also asymmetrically encrypted and stored in the database, so that it can be decrypted upon authorization by the institution's advisory board, if needed. The decryption key is store securely off-cloud to prevent accidental breaches.

The \emph{add contact} API is most frequently invoked, once every $15~mins$ by potentially $1000$'s of users. To avoid the power, compute, and network overheads of de/seria\-li\-zing JSON, we use an alternative binary format. It starts with $16~bytes$ of the \emph{source device ID}, which is followed by a series of \emph{scan records}, one per scan. Each record starts with a $4~bytes$ of UNIX \emph{epoch time} in seconds with the scan record's timestamp. The next $1~byte$ indicates the \emph{number of device contacts} `$n$' in that scan, followed by $17 \times n~bytes$ having the $16~byte$ \emph{device ID} and $1~byte$ \emph{RSSI value} for the $n$ proximate devices. If more than $n=255$ devices are found in one scan, the app creates multiple scan records. Records are created and sent by the app even if there are no proximate devices, since this information is also useful. As mentioned before, beacons are also encoded as device IDs following a standard UUID template. 

Intuitively, each record forms an adjacency list for the contact graph. The binary records from service calls from all users are appended to a file and every $2~hours$, a pre-processing service fetches these binary files and generates a corresponding CSV file with an edge list consisting of the timestamp, source device ID, sink device ID, and RSSI. This CSV file is backed up to Azure BLOB store and, as discussed later, stored on HDFS for further analytics.

\emph{Add GPS} is the next frequently called API, every 5~mins, for users who choose to share their GPS location. 
This data is used to generate a device density heatmap of the user's neighborhood for the recent past, and potentially for contact tracing. To support such spatio-temporal queries, we use the \emph{InfluxDB} temporal database to store the GPS data. One copy of the latitude and longitude is asymmetrically encrypted and stored in InfluxDB, along with the timestamp, to support authorized contact tracing. Another copy is transformed using a \emph{\Gls{GeoHash}}~\cite{morton1966computer} of $7$ characters, which reduces the precision of the location to a $150m \times 150m$ grid. When generating the heatmap for the app user's current location, we query over this GeoCode.

The app communicates hourly device health data using the \emph{add liveliness} API, as a set of key--value pairs that has evolved over app versions. As a result, we store this data within \emph{Azure Cosmos DB}, which is a NoSQL database. This data is later queried for identifying devices that are not reporting Bluetooth data reliably and to send them alerts with possible fixes, and also for monitoring the overall status of the GCG deployment at an institution.

Alerts are sent to the app using a custom \emph{notification service} in the backend that the app polls every $5~mins$. This approach was initially chosen over Google or Apple's \emph{push notifications} to reduce the dependence on external services. Alerts that are generated by various analytics are inserted into a MariaDB table with the device ID, title, content, type, and validity time range. When an app polls the service, any pending alerts for that device are returned. Besides displaying alerts to the user, they may also have a special payload that triggers changes to the UI, such as updating the social distancing score on the main screen. 

User-level analytics such as displaying their contact network and other analytics such as the user density are sent to the app as HTML that is locally rendered. The app invokes a \emph{get analytics} API which returns a JSON containing a list of current endpoints that serve the analytics. The plots and maps are served off an Apache 2 instance. Separately, we also run our own Open Street Maps \emph{tileserver} for serving the map tiles.

These external-facing services are hosted on a separate set of VMs over which the services are distributed based on their workload and to avoid performance interference. These VMs are shown in orange in Figure~\ref{fig:arch}.
We use one Azure D2s v3 VMs to host the \emph{register device, add gps}, and \emph{add liveliness} endpoints, a second one that exclusively runs the \emph{add contact} and another to run the \emph{get notifications} service; the latter two see a higher load. The \emph{tileserver} for displaying open street maps, which is only occasionally used, runs off an Azure B2S VM, while the analytics are served from an Azure D2s v3 VM. A separate Azure D4s v3 VM hosts MariaDB and InfluxDB used by these services.

\subsubsection{Internal services}
Besides the Internet-facing services, there are internal services to support the GCG platform. These are used to host an \emph{operations portal} to oversee the health of the system, on-boarding of devices, and visualize the contact network. The portal does not directly access any user database or files to prevent accidental access to or modifications of the raw data. Instead, a separate \emph{routing service} offers a limited set of well-defined services to access authorized data. These APIs are periodically called and the results are cached in a separate MariaDB instance used by the portal. The portal and its database are also hosted on separate VMs, shown in yellow in Figure~\ref{fig:arch}. This sandboxing also extends to the analytics services, which too do not directly access the user databases for sending alerts or generating visualizations, but operate through this routing API.

For example, the liveliness data is fetched every $15~mins$ through this routing service from Cosmos DB and into MariaDB for the portal to visualize the number of scan records received and scans failed among the apps, while the device registration summary is fetched through the API to plot the users on-boarded over time, distribution of their device make and models, etc.

\subsubsection{Securing the backend platform}
Ensuring the security of the services and the data collected by the GCG platform is of paramount importance, and is intrinsic to various design and deployment choices.
All the REST endpoints use HTTP/2 with \emph{HTTP Strict Transport Security (HSTS)}, which forces the use of a Transport Layer Security (TLS 1.2/SSL) encrypted channel between the GCG App and the backend, and prevents man-in-the-middle attacks.

\begin{figure}[t]
	\centering
	\includegraphics[width=1.0\textwidth]{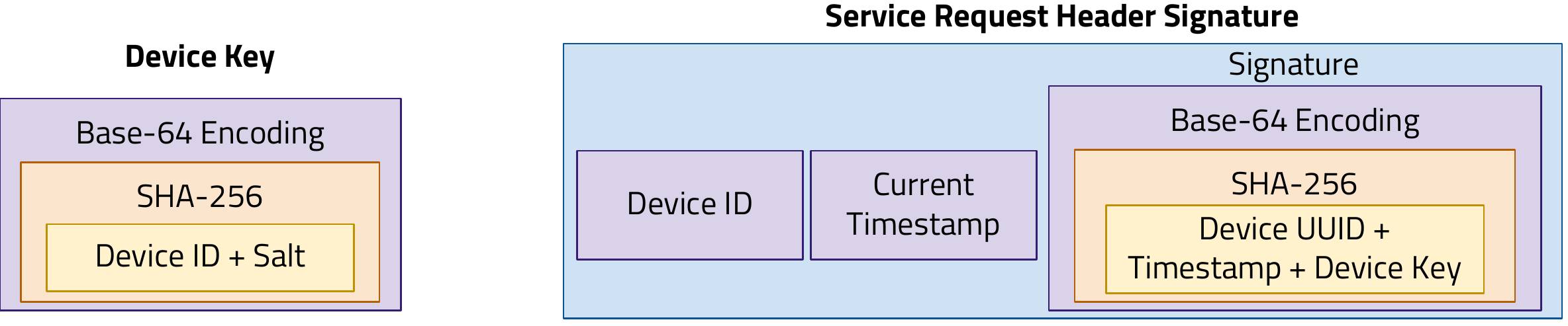}
	\caption{Signing service requests using device key}
	\label{fig:nonce}
\end{figure}

Further, all service calls are authenticated based on a \emph{device key} that is returned to the app during registration. To ensure that this service call authentication is light-weight, we use a \emph{digital signature protocol}, which ensures that each call can be locally validated, without the need for any database (Figure~\ref{fig:nonce}). Specifically, the device key is generated by the backend service as \texttt{key = base64(SHA256(device~ID, salt))}, where \texttt{salt} is a secret phrase known only to the service.
The GCG App encrypts and stores this device key on the phone.
Subsequently, when invoking any backend service, the app sends its device key, the current timestamp, and a signature, which consists of \texttt{sign = base64(SHA256(device~ID, timestamp, device~key))} as part of its HTTPS header or body. The service then uses the received device ID to generate the device key on the fly, and additionally uses the timestamp to generate the signature. It also verifies if the timestamp passed is recent, for mitigating replay attacks. If the generated signature matches the received signature, the request is valid and is executed. Note that all of these are flowing over an encrypted HTTPS channel.

Various other best security practises are used. The register device service takes measures to mitigate brute-force attacks using random invitation codes and PINs by limiting the number of daily attempts. Internal services such as the portal are only accessible from the institution's private network, over VPN, and are additionally secured using authentication. Firewall rules are used to restrict access to unused ports. Direct SSH access is not available to any VMs running services or the database. The Internet-facing VMs are in a separate subnet from the ones hosting the databases and internal services on Azure to keep the networks in different security domains. Data flows between the services and databases/storage is tightly controlled, and a routing service used for internal services.
We run the latest stable release of all software and the latest security patches to protect against known security flaws.

The MariaDB SQL database follows the principle of least privileges for access, and only minimal permissions for \texttt{SELECT} or \texttt{SELECT/INSERT} are given to user accounts. User-defined functions are disabled. All queries are templatized to avoid SQL code injection.
Sensitive data such as phone number and location are kept hashed and/or encrypted when stored. This prevents privacy from being compromised even if there is a cloud security breach and the data is leaked. We use asymmetric public-private keys so that only public keys are hosted on the VM for encryption and private keys for decryption are kept securely offline.
Contact data is backed up to Azure encrypted BLOB storage.

The backend services have undergone professional vulnerability and penetration testing by Crossbow Labs.\footnote{Crossbow Labs, \url{https://crossbowlabs.com/}}

\section{GCG Analytics and Contact Tracing}
\label{sec:analytics}
The GCG App is designed to provide feedback to users on their daily interactions using simple metrics and contact neighborhoods. Additionally, to improve user engagement, the app also provides heatmaps of user density and charts and maps that show the COVID-19 situation in various states and districts around the country. In this section, we describe these features along with the contact tracing protocols that are in place if an app user tests positive. 

\subsection{Temporal Network Analytics}
\subsubsection{Creating temporal graphs}
We receive contact records from various devices that contain the contact timestamp and associated Bluetooth signal value. For efficient primary and secondary contact tracing, we periodically stitch these contact records to create a global contact network graph. Further, we annotate the edges with the contact timestamps and signal values to creating a \emph{temporal contact network} or a \emph{\Gls{TemporalGraph}}. 

We use Apache Spark to perform this stitching from the CSV edge file, as a pre-processing step. Specifically, we create an \emph{interval graph} for scans received during a specific time interval. The Spark application takes a start and end time for the interval, and then filters in all the edge list entries in the input CSV file whose timestamp falls within this time interval. It then groups all edges by their source and sink vertices to create an adjacency list for each vertex that includes all scan entries from either source or sink edges. Every edge is characterised by a time interval $[t_s, t_e)$, where $t_s$ is the earliest scan timestamp and $t_e$ is the latest scan timestamp between the connecting devices, during that interval. Scans on an edge that fall on adjacent time points with the same RSSI value are combined to form longer intervals on the edge annotations. This gives a set of disjoint sub-intervals on the edge with an associated Bluetooth signal strength. The output is stored in HDFS for future analysis.

\begin{figure}[t]
	\centerline{\includegraphics[width=1\textwidth]{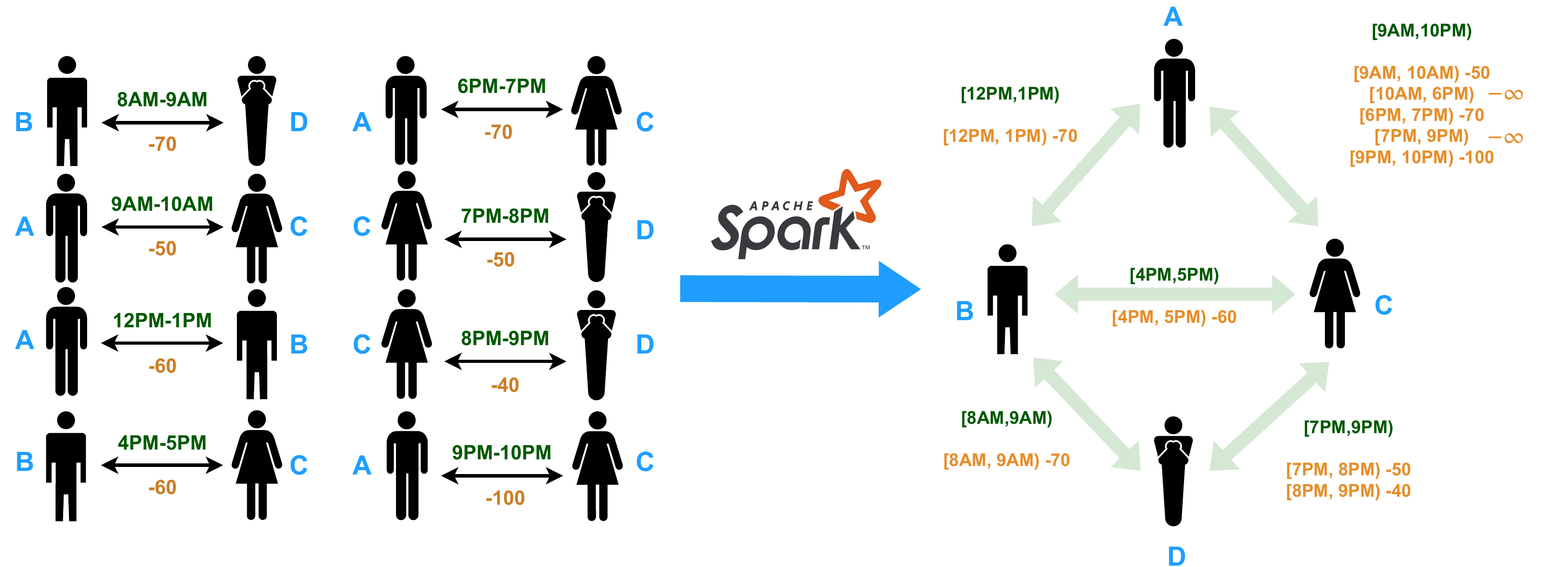}}
	\caption{Sample scan records for an interval and the corresponding interval graph}
	\label{fig:temporalgraph}
\end{figure}

Figure \ref{fig:temporalgraph} is an example interval graph obtained for a $24~hour$ time period. The interval has scan records for $4$ devices $A$--$D$, for $8$ contiguous time periods. Consider devices $A$ and $C$, which come in contact $3$ times with the earliest contact time being $9~AM$ and the last contact time being $10~PM$. The edge between $A$ and $C$ will thus span the time interval $[9~AM, 10~PM)$. This is further broken into sub-intervals: $[9~AM, 10~AM)$, $[10~AM, 6~PM)$, $[6~PM, 7~PM)$, $[7~PM, 9~PM)$, and $[9~PM, 10~PM)$, with corresponding signal strengths of $-80$, $-\infty$, $-70$, $-\infty$, and $-100$ respectively. A signal of $-\infty$ means the devices could not see each other, such as devices $A$ and $C$ between $[10~AM, 6~PM)$ and $[7~PM, 9~PM)$.

\subsubsection{Social Distancing scores}
The social distancing score provides users with a measure of their extent of social distancing, on a daily basis. Unlike the local Bluetooth data used to plot the contact counts on an hourly basis within the app, the social distancing score uses more global knowledge from a device and its neighbors. In particular, it accounts for ``background devices'' that are often or always in the vicinity, such as family members or hostel room neighbors, and which are subtracted from this score as their sustained presence does not pose any additional risk.

These scores are calculated using Apache Giraph once a day, over the interval graph created for the preceding $24~hour$ period. The score calculation depends on three parameters: signal threshold $(\delta)$, minimum contact duration $(\phi_m)$, and background contact duration $(\phi_b)$. 
For each device ID, we first identify those neighboring devices that could detect each other for at least $\phi_b~mins$, cumulatively, during the $24~hour$ period. These neighbors form the background devices and are eliminated from further analysis. Currently, we use $\phi_b = 240~mins$.

Next, from the remaining neighbors, we retain only the RSSI entries which exceed a value of $\delta$ on their edge sub-intervals. This helps identify the duration of \emph{nearby contacts} with them. Based on experiments described in the next section, we set $\delta=-78$, which approximates a distance of $2~m$.
We sum up the duration of nearby contacts for each edge, and those whose duration is greater than $\phi_m~mins$ form the proximate contacts,  $p$. We set $\phi_m = 15~mins$ by default. Intuitively, this means that the user has interacted with $p$ other devices in close physical proximity of about $\leq 2~m$ for a cumulative of $15~mins$ or more in the past $24~hours$, but who are not part of the sustained background presence.
From this, the social distancing score for a device is calculated as $\max\{0, 10-p\}$. This normalization offers a higher score for users who practise social distancing and a lower score for the others.

In the example snapshot, assume that $\delta=-60$, $\phi_m=30~mins$ and $\phi_b=180~min$. For the device $C$, devices $B$ and $D$ are proximate contacts since their close contact durations are $1~hr$ and $2~hrs$, respectively. However, $A$ is not a proximate neighbour of $C$ since it is a part of its background, having been detected for a total of $3~hrs$. So the social distancing score of $C$ is $8$. 

\subsection{Translating RSSI to Distance Measures}
The SARS-CoV-2 virus is currently assumed to spread by `contact and droplet' as well as airborne transmission ~\cite{coronaWho}. WHO and various countries have provided social distancing advisories that emphasize a minimum spacing of $1$--$2~m$ for curbing the spread of the virus~\cite{coronaWho, CDCsocialdist, countrywisesocialdist, indiasocialdist, chu2020}.
Being able to nudge users to maintain such distancing is one of the goals of the GCG App.

However, inferring distances accurately from Bluetooth RSSI values is non-trivial. Factors such as smartphone hardware variations, body interference, and multi-path interference lead to both false-positives and false-negatives while estimating the distance from RSSI values \cite{dyoung2020,dehaye2020}.

Researchers elsewhere have conducted experiments to understand if contact tracing apps can estimate if two users are close to each other, i.e., within a distance of $2~m$ for $15~mins$ or longer~\cite{Douglas2020}. These were performed with Google Pixel 2 and Samsung Galaxy A10 devices using the OpenTrace App\footnote{OpenTrace, \url{https://github.com/opentrace-community}}, an open-source version of Singapore's TraceTogether App~\cite{tracetogether}. They used different environmental conditions such as signal attenuation by the human body, a handbag, walls, etc. and also by enacting real-world scenarios.
The measured RSSI and the distance are plotted over time to understand the variability for different configurations and their relationship to the ground truth.

Another Smart Contract Tracing (SCT) System~\cite{ngpc2020} uses machine learning classifiers to classify the contacts as high/low risk using the Bluetooth RSSI values. They perform experiments to collect RSSI from a Nokia 8.1 with Android 10 and HTC M9 with Android 7.0 for distances ranging from $0.2$--$5~m$, and for random device orientations, and at different locations such as hand, pocket, and backpack. 
The collected data is labelled as +1 (high-risk, $\leq 2~m$) or -1 (low-risk) according to the ground truth.
They filter the data using a moving average filter before training using machine learning classifiers like decision tree, linear discriminant analysis, na\"{i}ve Bayes, $k$ nearest neighbors, and support vector machine.

The Google-Apple Exposure Notification API in Android also applies BLE calibration corrections based on manual measurement of the signal strength under standard conditions~\footnote{Exposure Notifications BLE RSSI calibration procedure, Google Developers, \url{https://developers.google.com/android/exposure-notifications/ble-attenuation-procedure}}.

Given the hardware diversity we observe among our campus population, we conduct similar lab-scale experiments, as described, using a more diverse number of smartphones and beacons. We evaluate the effect of RSSI at $1, 2$, and $4~m$ distances to help us determine whether two phones are within $2~m$.

\subsubsection{Experiment Design}
We use a debug version of the GoCoronaGo Android and iOS apps that log the Bluetooth scan information to a local file on the smartphone in our experiments. The experiment was performed in an open room measuring about $5 \times 5~m$ with few furniture, mimicking a real-world environment.  
Our experiment uses $9$ Android devices, $2$ iPhones, and $3$ Bluetooth Low Energy (BLE) beacons running both stock and custom Android OSes: Motorola Moto G6 (Android v9), two Motorola Moto G5S Plus (v8.1), Xiaomi Mi A3 (v9), Xiaomi Redmi Note 8 (Android v9, MIUI v11.0.2), Samsung Galaxy M31 (Android v10, One UI Core 2.0), Samsung Galaxy S9+ (Android v10, One UI 2.1), OPPO A1K (Android v9, ColorOS v6.0.1), and Vivo Y91i (Android v8.1, Funtouch OS v4.5), iPhone 7 (iOS v13.5) and iPhone XR (iOS v13.6). The BLE beacons from TechoLabz use the iBeacon protocol and transmit at $-3~dBm$ at $1~sec$ advertising intervals.

\begin{figure}%
    \centering
        \subfloat[$1~m$ distance]{{\includegraphics[width=0.5\textwidth]{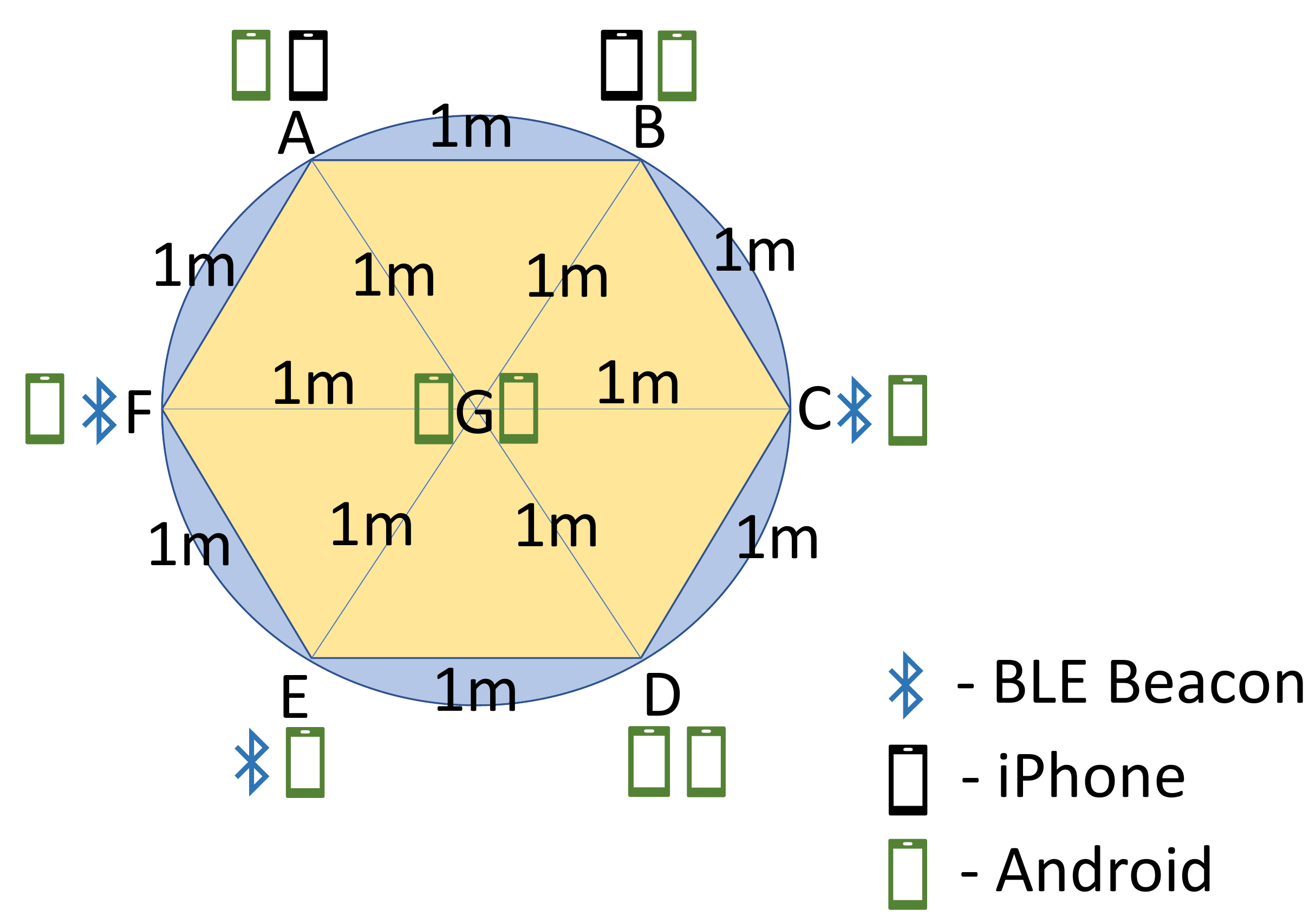}}%
    \label{fig:exp:rssi:1m}
    }%
    \qquad
    \subfloat[$2~m$ and $4~m$ distances]{{\includegraphics[width=0.4\textwidth]{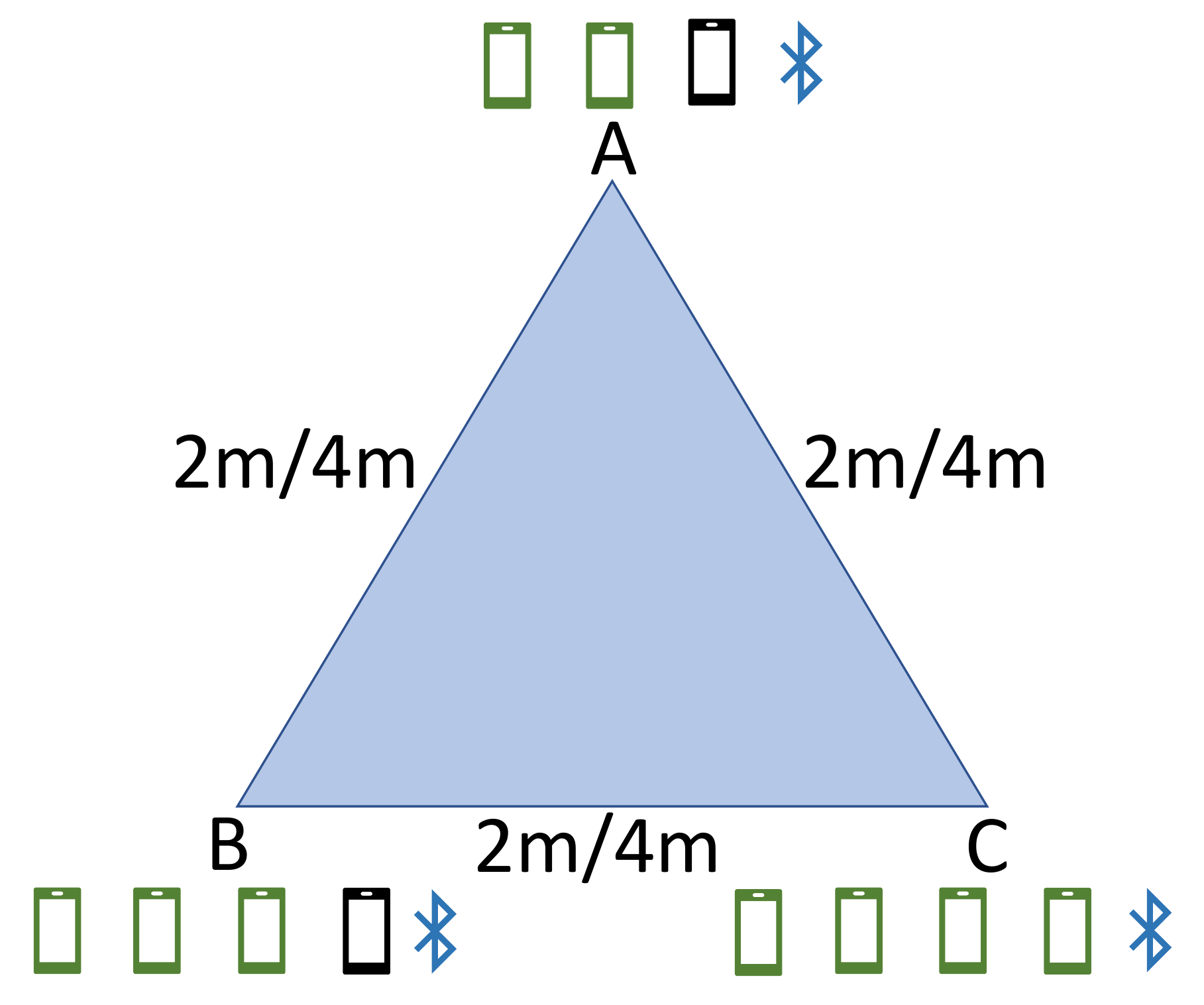}}%
    \label{fig:exp:rssi:2m4m}
    }%
    \caption{Experimental setup for collecting RSSI values at fixed distances}%
    \label{fig:exp:rssi}%
\end{figure}

All the devices were used at a high battery level, with power-saving modes disabled and screen set to stay on for as long as possible while performing the Bluetooth scans.
Each experiment configuration was performed for a period of $10~mins$ to give $\approx 10$ RSSI measurements per device pair in that configuration.
Given the technical limitations of iOS, Android devices can detect other Android devices and the Beacons, and iPhones can detect the Android devices. Considering these factors, two experimental setups were designed to collect the RSSI data as illustrated in Figure~\ref{fig:exp:rssi}.

For the distance $a = 1~m$, we use a hexagonal placement, as shown in Figure~\ref{fig:exp:rssi:1m}, with pairs of devices at the vertices, A, B, C, D, E, F, and the center, G. 
These give us devices at distances of $0~m$ (same vertex); $1~m$, between adjacent vertices, e.g., A--B; $2~m$, between vertices at diagonal corners, e.g., A--D; and $\sqrt{3}~m$ for vertices that are 2 hops away, e.g., A--C.
Three runs with the hexagonal setup are required to ensure that every pair of devices is measured at a $1~m$ distance.

For distances $a = 2~m$ and $4~m$ the devices were arranged in 3 clusters, A, B, C, at the corners on an equilateral triangle with a side of length $a$ (Figure ~\ref{fig:exp:rssi:2m4m}. 
In each cluster, the devices are placed vertically and adjacent to each other, in a row. Devices across clusters are separated by a distance $a$ while those within a cluster have a distance of $\approx 0~m$. Three runs of the triangular setup with different clusters are performed to ensure that we get the RSSI for each pair of devices at $2~m$ and $4~m$.

\subsubsection{Modeling proximity distances using RSSI values}

A key rationale for this study is to understand if two devices are within $2~m$ of each other or not, as we use the $2~m$ distance as the proximity threshold in our platform. A total of $1988$ RSSI data points at $1~m$, $2865$ data points at $2~m$, and $2321$ data points for $4~m$ are collected. We focus our analysis on just the Android phones, which form the bulk of our deployment. There are $1073$, $1746$, and $1377$ data points for $1,2$, and $4~m$ between the Android devices, respectively. For each distance and a device pair, we drop the maximum and minimum RSSI values to eliminate outliers.

\begin{figure}%
    \centering
    \subfloat[Empirical CDF of RSSI values at $1,2$ and $4~m$]{{\includegraphics[width=5cm]{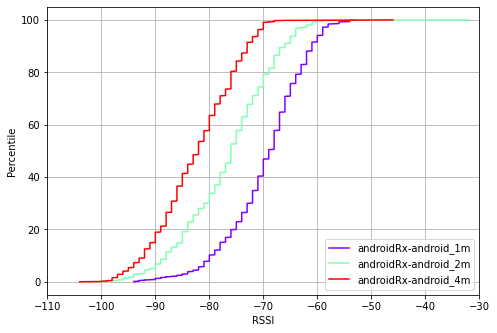}}%
    \label{fig:exp:rssi:plots:cdf}}
    \qquad
    \subfloat[Difference in Percentiles for RSSI at $2~m$ and $4~m$]{{\includegraphics[width=5cm]{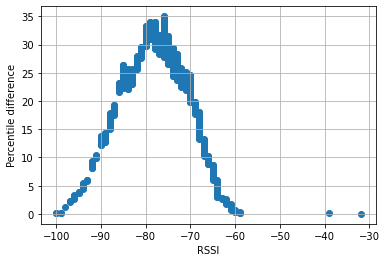}}%
    \label{fig:exp:rssi:plots:2m4m}}%
    \caption{Distribution of RSSI at different distances for the Android phones}%
    \label{fig:exp:rssi:plots}%
\end{figure}

An empirical Cumulative Distribution Function (CDF) of the RSSI values at $1,2$, and $4~m$ are shown in Figure~\ref{fig:exp:rssi:plots:cdf}. The X-axis shows the RSSI values while the Y-axis lists the corresponding percentiles for different distance configurations. We see that there is a substantial overlap between data points at the 3 different distances for a given RSSI. E.g., for an RSSI of $\leq -75$, we have $23\%$ of the $1~m$ data points, $54\%$ of the $2~m$ data points and $84\%$ of the $4~m$ data points fall within that signal strength.
So, using any single threshold value of RSSI as an estimate for a $2~m$ distance is liable to result in both false positives and false negatives.

For this preliminary study, we wish to determine an RSSI value that is the most discriminating with regard to the $\leq 2~m$ and $>2~m$ proximity. So for each RSSI value, we plot the difference in the percentile of data points that are at $2~m$ and at $4~m$ distances, and this is shown in Figure~\ref{fig:exp:rssi:plots:2m4m}. The peak difference is observed at an RSSI value of $-78$, i.e., the difference between the true positive of $2~m$ (59\%) and false positive of $4~m$ (29\%) is the highest. 
Hence, we use an RSSI of $-78$ as the proximity threshold in our GCG app and the backend analytics.

In the future, we propose to study the effect on RSSI from different pairs of phone models and in different environmental conditions in order to develop a more customized proximity threshold, instead of using a single global value that is currently adopted.

\subsection{Contact Tracing Protocol}
\label{sec:tracing}

When an app user tests positive for COVID or is under mandatory quarantine, the current protocol at IISc requires the campus health center to check if the user is willing to share their contact data for tracing. If so, they are asked to enter their phone number within the GCG App, if not done so.
The health center collects and enters the GCG unique id, device id suffix, and their phone number from the user into a portal. This initiates a call to the GCG backend and triggers an OTP to the user's phone number, if the details match with an existing user. The user may share this OTP with the health center and this serves as their \emph{informed consent for contact tracing}.

The health center enters the OTP and any additional details about the subject such as symptoms, start and end dates for contact tracing, and test information. The GCG backend confirms if the OTP is accurate, and if so, the request is forwarded to the advisory board to get the primary and secondary contacts for this user. The advisory board has representatives from the institute, including faculty, staff, students, doctors, and a bio-ethicist. 

If the board approves the request through their portal, the GCG backend is notified and it will perform a \emph{time-respecting breadth first search (T-BFS)}, which is a variant of \Gls{BFS} performed over the temporal contact graph. The T-BFS will be initiated from the device ID corresponding to the given user's unique ID and for the time duration in the past indicated by the health center. If the user's unique ID is associated with multiple devices during this period, the search will be initiated from each of these IDs. The output is a list of device ids for the primary and secondary contacts. We then use the invitation code, unique ID and device ID mappings maintained in the GCG backend to get the list of invitation codes used by the primary and secondary contacts.

These invitation codes are shared with the IT staff who then use their mapping table to de-anonymize them and provide the health center with a list of email IDs and/or phone numbers of these contacts. The GCG backend also provides the duration of contacts for each of the invite codes. The health center can then choose to initiate their relevant protocols for reaching out to these contacts, and quarantine or test them. If mandated by law, the health center may share the contact trace data with the local government agency responsible for COVID-19 surveillance.

\subsection{Other Analytics and User Engagement}
Besides the local analytics within the app, we also provide additional analytics to the GCG user based on aggregation in the backend.

Figure~\ref{fig:ui:density} shows a heatmap of GCG user count in a $1.5 \times 1.5~km$ area around the current location of an app user, if they share their GPS location. It is aggregated over the last $24~hours$ from users who share their GPS data. This data is queried from the timestamp and geohashes present in the InfluxDB backend. In order to \emph{respect privacy}, the location data is spatially coarsened into tiles of approximately $150~m \times 150~m$ area, and temporally coarsened over $24~hours$, and only the aggregate count of users in each tile is shown. Also, when few users are present in a tile, we display this data in a categorical manner, e.g., $<5$.

\begin{figure}
    \centering
    \includegraphics[scale=0.5]{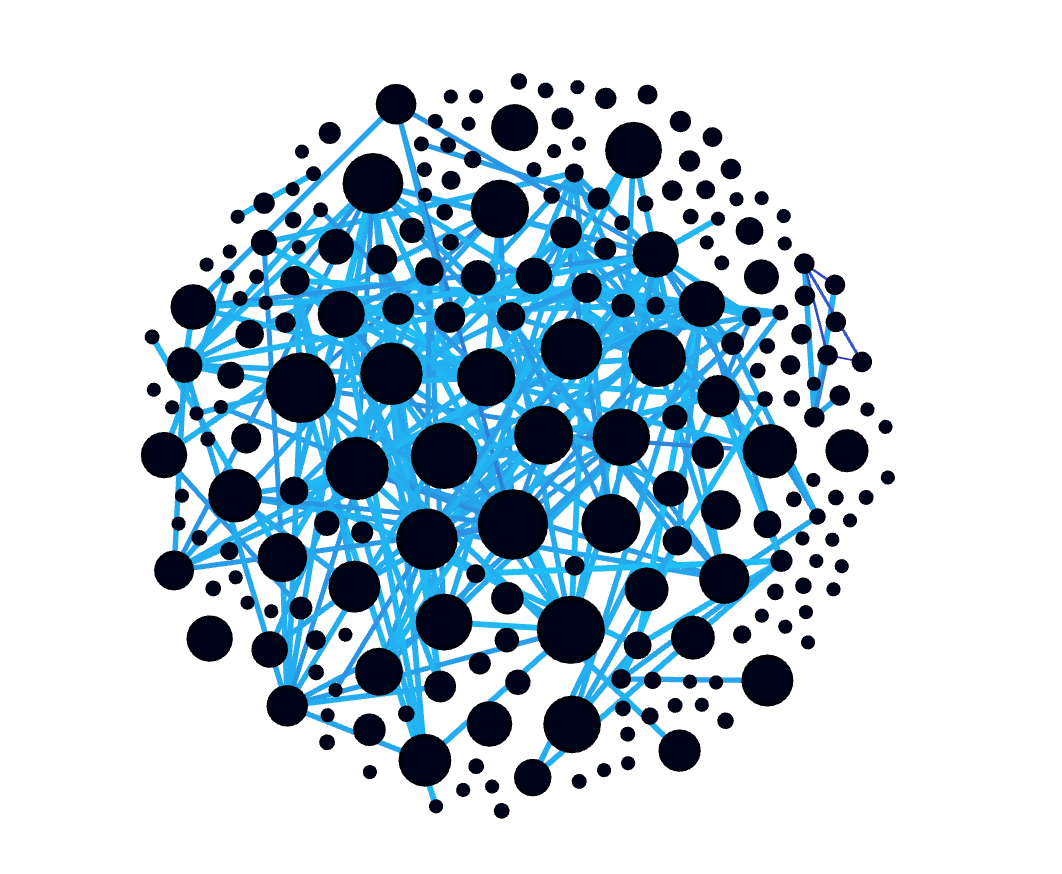}
    \caption{A visualization of the contact graph of a subset of app users for a single day. The size of the nodes is proportional to their degree centrality.}
    \label{fig:networkVisualisation}
\end{figure}

The contact graphs that are constructed in the backend can be visualized using tools such as Gephi. Figure~\ref{fig:networkVisualisation} shows a subset of the temporal graph generated for a single day. Here, the size of a node depends on its degree \Gls{CentralityMeasure} across the entire time duration. The thickness of the links depends on the duration of their contact.

While such a graph is instructive for backend analytics, we use it to generate a \emph{neighbourhood tree} for each user, as shown in Figure~\ref{fig:ui:nw}. The tree is based on the last $48~hours$ of data and contains contacts up to 2-hops. Importantly, this is a tree and not a neighborhood sub-graph to \emph{preserve privacy}, i.e., edges between the 1-hop and 2-hop neighbors are not shown to avoid revealing contact patterns between them. These trees are generated on a daily basis. It helps the users get a sense of not just their primary contacts, but also their secondary contacts, which could be much larger, and in-turn motivate users to take greater precautions by socially distancing. 

\ysnoted{Later, JV to add ops portal}

\section{Discussion}
\label{sec:discussion}

\begin{figure}[t]
\centering
\includegraphics[width=1\textwidth]{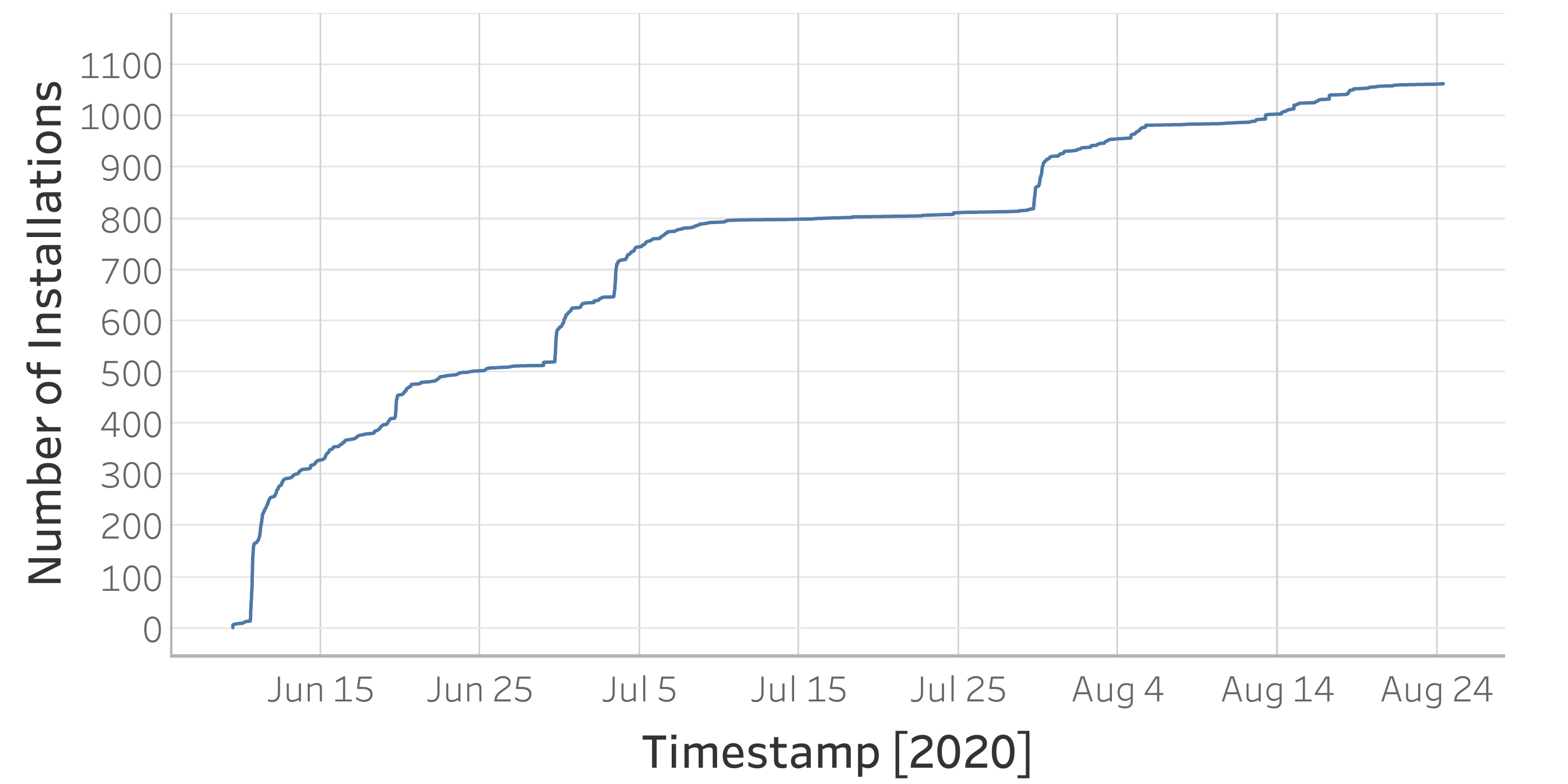}
\caption{Number of GCG installations at IISc over time}
\label{fig:installations}
\end{figure}

\subsection{Deployment Experience}
The GCG App is currently deployed at the Indian Institute of Science (IISc), Bangalore. The IISc campus is an access controlled residential campus with close to 4,000 students, over 450 faculty, and over 2,000 research and administrative staff. A majority of the students and faculty live on campus. However, IISc entered a full shutdown in March, 2020, a few days ahead of a nation-wide lockdown in India, and the students on campus were instructed to leave for their homes. Initial versions of the app were tested among faculty volunteers during the lockdown period. The GCG App was first rolled out to students in June, 2020 after a subset of them were allowed to re-enter campus, and subsequently to other faculty and staff.

At the time of writing this paper, the GCG App has been installed by over 1,000 users at IISc. A plot of the number of installations of the GCG App over time is shown in Figure \ref{fig:installations}. Sharp jumps in installations correspond to new invitations or reminders sent to students, faculty, and staff for installing the app. 
The app is yet to be rolled out to essential workers such as hostel cooks, cleaning staff and security personnel, and noticeably, some of the early cases of COVID-19 on campus have been initiated through them. This is understandable since many of them stay off-campus and possibly have a larger mobility footprint, increasing their risk of acquiring the coronavirus.

While the GCG Android App was initially hosted on the IISc website due to restrictions by Google and Apple in hosting COVID-related apps on their online app stores, it has recently received approval to be hosted on the Google Play Store, with v0.5 currently available there since early August, 2020. An \emph{ad hoc} iOS version is also being tested since the last week of August, 2020.

\subsection{Federated Deployment}
While GCG is designed for institutional use, contact tracing for users from the same institutions who interact outside the campus is also captured. This benefit can be further expanded through a federated deployment for institutions that are spatially close to each other, such as a cluster of college campuses and software tech-parks in the same neighborhood. Here, the chances of physical interaction between users from different organizations are high, e.g., visiting the same local cafeteria or grocery store.

\begin{figure}[t]
    \centering
    \includegraphics[width=0.8\textwidth]{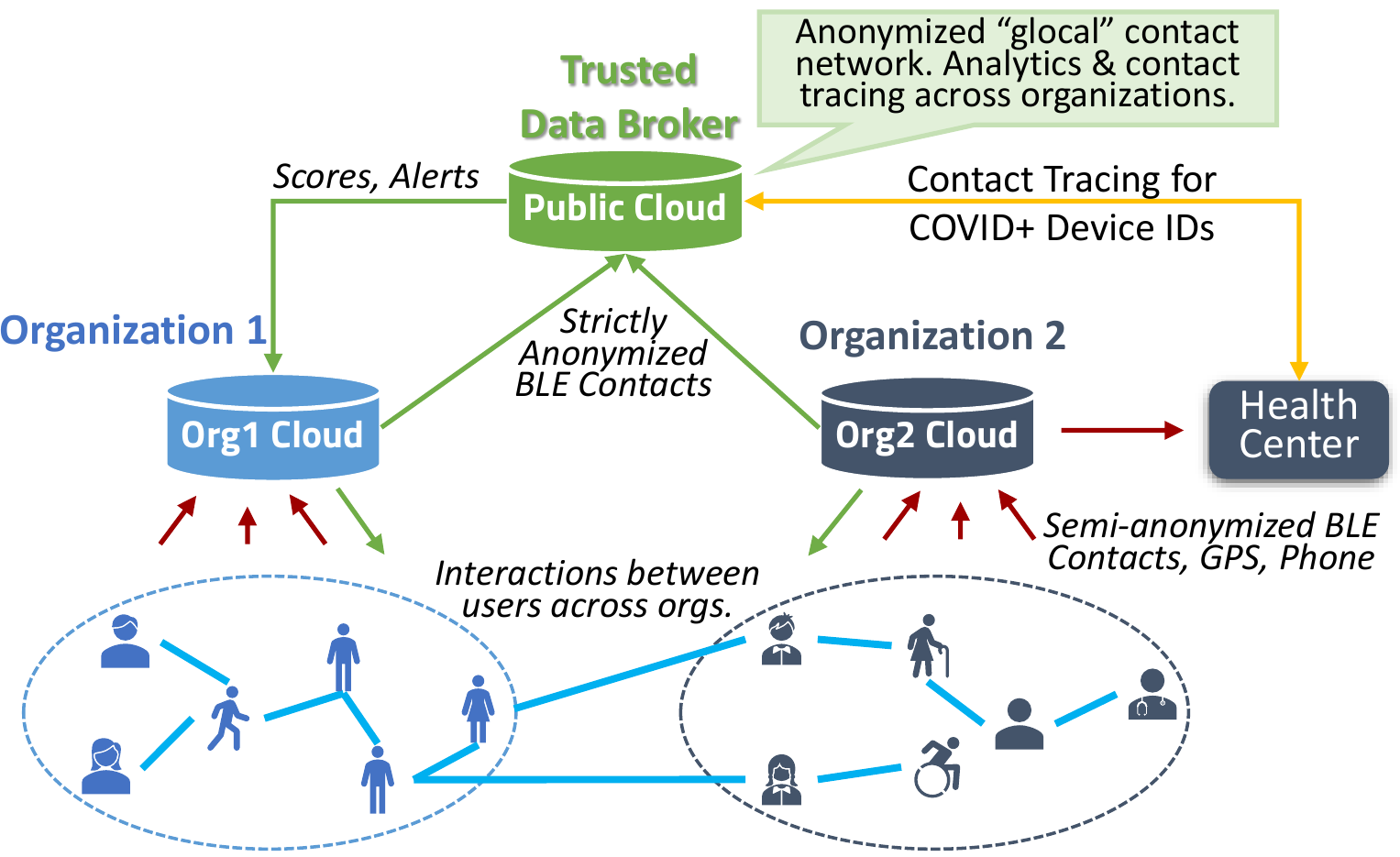}
    \caption{Federated deployment of GCG across Institutions}
    \label{fig:fed}
\end{figure}

In this federated deployment, individual institutions would maintain their independent GCG deployments. But in addition, they would share the strictly anonymized contact graph for their institution with a \emph{trusted data broker}, such as a non-profit agency or a neutral university. This data broker would then stitch these graphs together based on contacts between unique device ids that span graphs from different institutions. This can then be used to trigger ``glocal'' analytics -- a global combination of local clusters that are near each other -- and share more accurate proximity scores with the users of individual institutions, as well as perform more effective contact tracing across institutions in the same community.
A key requirement to preserving privacy is that no personal data should be shared with this trusted broker, and any de-anonymization for contact tracing should strictly be handled at the local institution.

This can further be complemented through the use of national or regional-scale contact tracing apps, even if used by a smaller fraction of users who are mobile. This can help link clusters of GCG contacts within institutions, and allow with contact tracing beyond the institutional premises as well. However, care should be taken to sandbox the regional and institutional datasets to avoid privacy loss.

\subsection{Research Opportunities}
The availability of fine-grained contact tracing data has opened opportunities for new research on infection spreading. Classic epidemiological models are compartmentalized formulations that classify the population into different states such as \textit{S (Susceptible)}, \textit{E (Exposed)}, \textit{I (infected)}, and \textit{R (removed/recovered)}. Based on the progression patterns of a disease, different models such as SI, SIS, SIR, and SEIR models~\cite{kermack27, anderson92, barabasi16, kiss17} have been proposed. These models are applicable to large populations and can estimate the time evolution of the fraction of individuals in different states over time, and can identify the peak number of infections for different \Gls{ReproductionNumber}s. The assumptions in these models are however coarse and their utility is hence limited. They can be used to take higher-level policy decisions such as deciding the duration of lockdowns, planning hospital bed-capacity over time, etc. However, the input data for these models is tightly related to the testing rates, which in the case of COVID-19 was very low during the initial few months. 

Research in the last two decades has extended such compartmentalized models to static or time-varying contact networks~\cite{wang03, van08, ganesh05, youssef11}. In a static network, a node, if infected, can potentially infect any other nodes that it comes in contact with, regardless of the time of contact. But in dynamic networks, temporal ordering is preserved. That is, if an individual A comes in contact with a person B before B and C interacted, then A faces no risk from C. This can correct for the over-prediction of infection rates from static models. With Bluetooth-based mobile contact tracing, it is possible to include both duration of contact and the signal strength, which is a proxy for the distance between the phone users during their interaction, to make better predictions of the transmission rates. Results from simulated experiments by Kretzschmar \textit{et al.}~\cite{kretzschmar2020impact}, indicate reduced reproduction numbers when contact tracing is performed using mobile apps as the delay in alerting vulnerable individuals is reduced to a minimum. Apart from identifying primary and higher-order contacts quickly, contact data allows us to identify the most vulnerable users through either simulations of network models assuming hypothetical initial conditions or centrality measures. Most centrality scores from network science are defined on static graphs and it would be interesting to develop better centrality measures that can be used to find the nodes with higher spreading capabilities in a temporal network. Identifying such individuals can in-turn be used to device adaptive testing and vaccination strategies, which can help improve the estimates of the health states of the population, especially when testing is expensive, or its availability is limited.

Another major opportunity with centralized contact tracing is the ability to influence social distancing behaviour using alerts and scores. Creating control groups and providing such information to one of them and observing their contact patterns for a limited subsequent period can throw light on the effect of such scores. Such randomized control trials can help quantify the effectiveness of contact tracing apps even in the absence of COVID-19 case data.

\subsection{Challenges}
One of the key challenges with digital contact tracing is user adoption. As highlighted in Section~\ref{sec:background}, digital contact tracing requires a large fraction of users within the community to use it before it becomes effective. Having only a small sample of individuals use the app makes it difficult to identify the true sources of infection, because of which paths between infected individuals and their primary and higher-order contacts may go undetected. 

However, our experience with institutional-level contact tracing appears more promising than that employed by governments at a national level in terms of the fraction of users installing an app and the duration for which they had it installed on their phones. In fact, recent reports indicate that even $15\%$ of user adoption of contact tracing apps can have a meaningful impact of $6$--$8\%$ reduction in COVID infections and death~\cite{Abueg2020.08.29.20184135}. That said, not all workplaces are captive environments. In such cases, neighborhood or regional deployments of contact tracing apps
may be required since they are more likely to interact with people outside their cluster. Further, people may also interact during activities outside workplaces and their institutional contact tracing app can be ineffective during these periods.

We frequently observe app users turn off their Bluetooth or GPS, because of which the contact trace data collected is curtailed. Users may do so to save battery -- even though our experience shows that the Android app consumes less than $10\%$ of batter in an entire day -- or when they perceive a lower risk based on their current activity and environmental conditions. These factors can dramatically offset the promises offered by network-based epidemiological models in identifying risk-prone individuals and in contact tracing to contain the spread of infection. It is also extremely difficult to impute such missing data and no assumption can be confidently justified.

Although digital contact tracing apps have several potential advantages, validating its usefulness is tough. The difference between the two approaches can be best demonstrated when there are COVID positive app users who have shared data for continuous periods. In practice, it is wise to use data from such tools in conjunction with manual contact tracing since there would be gaps in data due to user behaviour or technology limitations. Building robust epidemiological models is all the more challenging because they contain several parameters that have to be calibrated from sparse and missing data. Heavy reliance on digital contact tracing apps can also exclude fractions of the community who use feature phones. Visitors to institutions such as delivery providers can also be missed out but can contribute to virus spreading.

Digital contact tracing is still in its infancy. It is important that individuals understand the data shared, risks, and benefits before fully using such apps. Communicating these details to a lay audience can be challenging and misconceptions about what such apps collect and can do are not uncommon.

\section{Conclusions}
In this article, we have described the various dimensions of digital contact tracing for managing the COVID-19 pandemic. We have highlighted the approaches taken by diverse apps globally, and their pros and cons. We have proposed GoCoronaGo as an institutional contact tracing app, whose design choices attempt to balance the privacy of individuals with the safety of the community in performing rapid multi-hop contact tracing. We have offered a detailed technical description of the GCG App, its backend services, and analytics. This platform is currently being validated at the IISc university campus, with additional campus deployments underway. We have shared our early experiences with the deployment over the past few months, in the midst of the COVID-19 epidemic, and the opportunities and challenges that lie ahead. Given the evolving nature of COVID-19, our continued experience with this contact tracing platform at IISc and other campuses can serve as a role model, or a cautionary tale, in managing the pandemic in the ensuing months and years.

\section*{Acknowledgements}
The authors acknowledge a research grant from the Department of Science and Technology (DST), Government of India, to partly sponsor this work (Grant No. DST/ICPS/\-RAKSHAK/2020). We also recognize the support offered by the RAKSHAK review committee. 
We thank the administration of IISc for assistance with the development and deployment of GCG, the members of the Institute who volunteered to test early versions of the app, and Prof. Y. Narahari who offered valuable guidance to the project. We are grateful for the detailed feedback offered by the Institute Human Ethics Committee (IHEC) at IISc in designing the operations and the research study. A proposal for research using the data collected from the app is currently under review by IHEC. 
We are also glad for valuable inputs from Dr. Olinda Timms from St. Johns Research Institute and Prof. Mukund Thattai from NCBS on the design of the contact tracing protocol to balance safety and privacy.
A special thanks to Crossbow Labs for their \emph{pro bono} security testing services.

\bibliographystyle{abbrv}
\bibliography{references}

\begin{thebibliography}{10}

\bibitem{indiasocialdist}
\enquote{Advisory on Social Distancing Measure in view of spread of COVID-19
  disease}.
\newblock Technical report, Ministry of Health and Familty Welfare, India,
  2020.

\bibitem{10665-332049}
\enquote{Contact tracing in the context of COVID-19: interim guidance}.
\newblock Technical report, {World Health Organization (WHO)}, 2020.

\bibitem{coronaWho}
\enquote{Coronavirus disease (COVID-19) advice for the public}.
\newblock Technical report, {World Health Organization (WHO)}, 2020.

\bibitem{bbcwrist}
\enquote{Coronavirus: People-tracking wristbands tested to enforce lockdown}.
\newblock {\em BBC}, 2020.
\newblock \url{https://www.bbc.com/news/technology-52409893}.

\bibitem{10665-332265}
\enquote{Digital tools for COVID-19 contact tracing: annex: contact tracing in
  the context of COVID-19}.
\newblock Technical report, {World Health Organization (WHO)}, 2020.

\bibitem{google-en}
\enquote{Exposure Notifications: Using technology to help public health
  authorities fight COVID‑19}.
\newblock Technical report, Google and Apple, 2020.
\newblock \url{https://www.google.com/covid19/exposurenotifications/}.

\bibitem{pepp}
\enquote{{Pan-European Privacy-Preserving Proximity Tracing (PEPP-PT)}}, 2020.
\newblock \url{https://www.pepp-pt.org/}.

\bibitem{CDCsocialdist}
\enquote{Social Distancing}.
\newblock Technical report, {Centers for Disease Control and Prevention (CDC)},
  2020.

\bibitem{nz-covid}
{\enquote{NZ COVID Tracer app}}, 2020.
\newblock
  \url{https://www.health.govt.nz/our-work/diseases-and-conditions/covid-19-novel-coronavirus/covid-19-novel-coronavirus-resources-and-tools/nz-covid-tracer-app}.

\bibitem{Abueg2020.08.29.20184135}
M.~Abueg, R.~Hinch, N.~Wu, L.~Liu, W.~J.~M. Probert, A.~Wu, P.~Eastham,
  Y.~Shafi, M.~Rosencrantz, M.~Dikovsky, Z.~Cheng, A.~Nurtay,
  L.~Abeler-D{\"o}rner, D.~G. Bonsall, M.~V. McConnell,
  S.~O{\textquoteright}Banion, and C.~Fraser.
\newblock Modeling the combined effect of digital exposure notification and
  non-pharmaceutical interventions on the covid-19 epidemic in washington
  state.
\newblock Technical report, {University of Oxford and Google}, 2020.

\bibitem{ahmed-ieee-surv}
N.~{Ahmed}, R.~A. {Michelin}, W.~{Xue}, S.~{Ruj}, R.~{Malaney}, S.~S.
  {Kanhere}, A.~{Seneviratne}, W.~{Hu}, H.~{Janicke}, and S.~K. {Jha}.
\newblock \enquote{A Survey of COVID-19 Contact Tracing Apps}.
\newblock {\em IEEE Access}, 8:134577--134601, 2020.

\bibitem{anderson92}
R.~M. Anderson, B.~Anderson, and R.~M. May.
\newblock \enquote{Infectious diseases of humans: dynamics and control}.
\newblock {\em Epidemiology and Infection}, 108(1):211, 1992.

\bibitem{backer20}
J.~A. Backer, D.~Klinkenberg, and J.~Wallinga.
\newblock \enquote{Incubation period of 2019 novel coronavirus (2019-nCoV)
  infections among travellers from Wuhan, China, 20--28 January 2020}.
\newblock {\em Eurosurveillance}, 25(5):2000062, 2020.

\bibitem{barabasi16}
A.-L. Barab{\'a}si et~al.
\newblock {\em \enquote{Network science}}.
\newblock Cambridge university press, 2016.

\bibitem{bassi2020overview}
A.~Bassi, S.~Arfin, O.~John, and V.~Jha.
\newblock \enquote{An overview of mobile applications (apps) to support the
  coronavirus disease-2019 response in India}.
\newblock {\em The Indian Journal of Medical Research}, 15(5), 2020.

\bibitem{bay2020bluetrace}
J.~Bay, J.~Kek, A.~Tan, C.~S. Hau, L.~Yongquan, J.~Tan, and T.~A. Quy.
\newblock \enquote{BlueTrace: A privacy-preserving protocol for
  community-driven contact tracing across borders}.
\newblock Technical report, Government Technology Agency-Singapore, 2020.

\bibitem{Bertuletti2016}
S.~{Bertuletti}, A.~{Cereatti}, M.~{Caldara}, M.~{Galizzi}, and U.~{Della
  Croce}.
\newblock \enquote{Indoor distance estimated from Bluetooth Low Energy signal
  strength: Comparison of regression models}.
\newblock In {\em IEEE Sensors Applications Symposium (SAS)}, pages 1--5, 2016.

\bibitem{btsig}
{Bluetooth SIG}.
\newblock \enquote{Bluetooth SIG to Extend Reach of COVID-19 Exposure
  Notification Systems}, 2020.
\newblock
  \url{https://www.bluetooth.com/press/bluetooth-sig-to-extend-reach-of-covid-19-exposure-notification-systems/}.

\bibitem{Braithwaite-lancet-surv}
I.~Braithwaite, T.~Callender, M.~Bullock, and R.~W. Aldridge.
\newblock \enquote{Automated and partly automated contact tracing: a systematic
  review to inform the control of COVID-19}.
\newblock {\em The Lancet Digital Health}, 2020.

\bibitem{wired-nhsx}
M.~Burgess.
\newblock \enquote{Why the NHS COVID-19 contact tracing app failed}.
\newblock {\em Wired Magazine}, June 2020.

\bibitem{sensortower}
S.~Chan.
\newblock \enquote{COVID-19 Contact Tracing Apps Reach 9\% Adoption In Most
  Populous Countries}, 2020.
\newblock \url{https://sensortower.com/blog/contact-tracing-app-adoption},
  Accessed on 20/07/2020.

\bibitem{chu2020}
D.~Chu et~al.
\newblock \enquote{Physical distancing, face masks, and eye protection to
  prevent person-to-person transmission of SARS-CoV-2 and COVID-19: a
  systematic review and meta-analysis}.
\newblock {\em The Lancet}, 395, 06 2020.

\bibitem{countrywisesocialdist}
H.~Coffey.
\newblock \enquote{SOCIAL DISTANCING: THE SCIENCE BEHIND REDUCING FROM TWO
  METRES TO ONE METRE}.
\newblock {\em Independent}, 2020.

\bibitem{covidsafe}
{Commonwealth of Australia}.
\newblock {\enquote{COVIDSafe App}}.
\newblock
  \url{https://www.health.gov.au/resources/apps-and-tools/covidsafe-app}.

\bibitem{tracetoken}
S.~CROSS.
\newblock \enquote{Trace Together Token: Teardown and Design Overview}, 2020.
\newblock \url{https://xobs.io/trace-together-token-teardown/}.

\bibitem{dehaye2020}
P.-O. Dehaye.
\newblock \enquote{Inferring distance from Bluetooth signal strength: a deep
  dive}, 2020.
\newblock
  \url{https://medium.com/personaldata-io/inferring-distance-from-bluetooth-signal-strength-a-deep-dive-fe7badc2bb6d},
  \text{Accessed on 18/07/2020}.

\bibitem{smh-aus}
{Editorial Board}.
\newblock \enquote{Much-hyped contact-tracing app a terrible failure}.
\newblock {\em The Sydney Morning Herald}, June 2020.

\bibitem{enright18}
J.~Enright and R.~R. Kao.
\newblock \enquote{Epidemics on dynamic networks}.
\newblock {\em Epidemics}, 24:88--97, 2018.

\bibitem{Ferrettieabb6936}
L.~Ferretti et~al.
\newblock \enquote{Quantifying SARS-CoV-2 transmission suggests epidemic
  control with digital contact tracing}.
\newblock {\em Science}, 368(6491), 2020.

\bibitem{ganesh05}
A.~Ganesh, L.~Massouli{\'e}, and D.~Towsley.
\newblock \enquote{The effect of network topology on the spread of epidemics}.
\newblock In {\em IEEE Annual Joint Conference of the IEEE Computer and
  Communications Societies}, volume~2, pages 1455--1466, 2005.

\bibitem{he20}
X.~He, E.~H. Lau, P.~Wu, X.~Deng, J.~Wang, X.~Hao, Y.~C. Lau, J.~Y. Wong,
  Y.~Guan, X.~Tan, et~al.
\newblock \enquote{Temporal dynamics in viral shedding and transmissibility of
  COVID-19}.
\newblock {\em Nature medicine}, 26(5):672--675, 2020.

\bibitem{guard-nhsx}
A.~Hern and D.~Sabbagh.
\newblock \enquote{Critical mass of Android users crucial for NHS
  contact-tracing app}.
\newblock {\em The Guardian}, May 2020.

\bibitem{hinch2020effective}
R.~Hinch, W.~Probert, A.~Nurtay, M.~Kendall, C.~Wymant, M.~Hall, and C.~Fraser.
\newblock \enquote{Effective configurations of a digital contact tracing app: A
  report to NHSX}.
\newblock Technical report, BDI Pathogen Dynamics team, University of Oxford,
  2020.

\bibitem{kao06}
R.~R. Kao, L.~Danon, D.~M. Green, and I.~Z. Kiss.
\newblock \enquote{Demographic structure and pathogen dynamics on the network
  of livestock movements in Great Britain}.
\newblock {\em Proceedings of the Royal Society B: Biological Sciences},
  273(1597):1999--2007, 2006.

\bibitem{kermack27}
W.~O. Kermack and A.~G. McKendrick.
\newblock \enquote{A contribution to the mathematical theory of epidemics}.
\newblock {\em Proceedings of the royal society of london. Series A, Containing
  papers of a mathematical and physical character}, 115(772):700--721, 1927.

\bibitem{kiss17}
I.~Z. Kiss, J.~C. Miller, P.~L. Simon, et~al.
\newblock \enquote{Mathematics of epidemics on networks}.
\newblock {\em Cham: Springer}, 598, 2017.

\bibitem{koher19}
A.~Koher, H.~H. Lentz, J.~P. Gleeson, and P.~H{\"o}vel.
\newblock \enquote{Contact-based model for epidemic spreading on temporal
  networks}.
\newblock {\em Physical Review X}, 9(3):031017, 2019.

\bibitem{kretzschmar2020impact}
M.~E. Kretzschmar, G.~Rozhnova, M.~C. Bootsma, M.~van Boven, J.~H. van~de
  Wijgert, and M.~J. Bonten.
\newblock \enquote{Impact of delays on effectiveness of contact tracing
  strategies for COVID-19: a modelling study}.
\newblock {\em The Lancet Public Health}, 5(8):e452--e459, 2020.

\bibitem{Kucharski2020lancet}
A.~J. Kucharski et~al.
\newblock \enquote{Effectiveness of isolation, testing, contact tracing, and
  physical distancing on reducing transmission of SARS-CoV-2 in different
  settings: a mathematical modelling study}.
\newblock {\em The Lancet Infectious Diseases}, 2020.

\bibitem{Douglas2020}
D.~J. Leith and S.~Farrell.
\newblock \enquote{Coronavirus Contact Tracing: Evaluating The Potential Of
  Using Bluetooth Received Signal Strength For Proximity Detection}.
\newblock Technical Report 2006.06822, arXiv, 2020.

\bibitem{li2020decentralized}
T.~Li, C.~Faklaris, J.~King, Y.~Agarwal, L.~Dabbish, J.~I. Hong, et~al.
\newblock \enquote{Decentralized is not risk-free: Understanding public
  perceptions of privacy-utility trade-offs in COVID-19 contact-tracing apps}.
\newblock Technical Report 2005.11957, arXiv, 2020.

\bibitem{liang2020covid}
L.-L. Liang, C.-H. Tseng, H.~J. Ho, and C.-Y. Wu.
\newblock \enquote{COVID-19 mortality is negatively associated with test number
  and government effectiveness}.
\newblock {\em Scientific reports}, 10(1):1--7, 2020.

\bibitem{novid}
P.-S. Loh.
\newblock \enquote{Accuracy of Bluetooth-Ultrasound Contact Tracing:
  Experimental Results from NOVID iOS Version 2.1 Using Five-Year-Old Phones}.
\newblock Technical report, 2020.

\bibitem{morton1966computer}
G.~M. Morton.
\newblock {\em \enquote{A computer oriented geodetic data base and a new
  technique in file sequencing}}.
\newblock International Business Machines Company New York, 1966.

\bibitem{asethu}
{National Informatics Centre}.
\newblock \enquote{Aarogya Setu}, 2020.
\newblock \url{https://www.mygov.in/aarogya-setu-app/}, Accessed on 24/07/2020.

\bibitem{ngpc2020}
P.~C. Ng, P.~Spachos, and K.~Plataniotis.
\newblock \enquote{COVID-19 and Your Smartphone: BLE-based Smart Contact
  Tracing}.
\newblock Technical Report 2005.13754, arXiv, 2020.

\bibitem{tracetogether}
G.~of~Singapore.
\newblock {\enquote{TraceTogether}}.
\newblock \url{https://www.tracetogether.gov.sg/}.

\bibitem{mitreview-adoption}
P.~H. O'Neill.
\newblock \enquote{No, coronavirus apps don’t need 60\% adoption to be
  effective}.
\newblock Technical report, MIT Technology Review, 2020.

\bibitem{ortiz06}
A.~Ortiz-Pelaez, D.~Pfeiffer, R.~Soares-Magalhaes, and F.~Guitian.
\newblock \enquote{Use of social network analysis to characterize the pattern
  of animal movements in the initial phases of the 2001 foot and mouth disease
  (FMD) epidemic in the UK}.
\newblock {\em Preventive veterinary medicine}, 76(1-2):40--55, 2006.

\bibitem{uber-cnn}
J.~Pagliery.
\newblock \enquote{Uber removes racy blog posts on prostitution, one-night
  stands}.
\newblock {\em CNN Business}, 2014.

\bibitem{rivest2020pact}
R.~L. Rivest, J.~Callas, R.~Canetti, K.~Esvelt, D.~K. Gillmor, Y.~T. Kalai,
  A.~Lysyanskaya, A.~Norige, R.~Raskar, A.~Shamir, et~al.
\newblock \enquote{The PACT protocol specification}.
\newblock {\em Private Automated Contact Tracing Team, MIT, Cambridge, MA, USA,
  Tech. Rep. 0.1}, 2020.

\bibitem{hrw-location}
K.~Roth.
\newblock \enquote{Mobile Location Data and COVID-19: Q\&A}.
\newblock Technical report, Human Rights Watch (HRW), 2020.

\bibitem{schafer16}
I.~J. Schafer, E.~Knudsen, L.~A. McNamara, S.~Agnihotri, P.~E. Rollin, and
  A.~Islam.
\newblock \enquote{The Epi Info Viral Hemorrhagic Fever (VHF) application: a
  resource for outbreak data management and contact tracing in the 2014--2016
  West Africa Ebola epidemic}.
\newblock {\em The Journal of infectious diseases}, 214(suppl\_3):S122--S136,
  2016.

\bibitem{megan-spectrum-2020}
M.~Scudellari.
\newblock \enquote{COVID-19 Digital contact tracing: Apple and Google work
  together as MIT tests validity}.
\newblock {\em IEEE Spectrum}, 13, 2020.

\bibitem{stehle11}
J.~Stehl{\'e}, N.~Voirin, A.~Barrat, C.~Cattuto, V.~Colizza, L.~Isella,
  C.~R{\'e}gis, J.-F. Pinton, N.~Khanafer, W.~Van~den Broeck, et~al.
\newblock \enquote{Simulation of an SEIR infectious disease model on the
  dynamic contact network of conference attendees}.
\newblock {\em BMC medicine}, 9(1):87, 2011.

\bibitem{dp3t}
C.~Troncoso, M.~Payer, J.-P. Hubaux, M.~Salathé, J.~Larus, E.~Bugnion,
  W.~Lueks, T.~Stadler, A.~Pyrgelis, D.~Antonioli, L.~Barman, S.~Chatel,
  K.~Paterson, S.~Čapkun, D.~Basin, J.~Beutel, D.~Jackson, M.~Roeschlin,
  P.~Leu, B.~Preneel, N.~Smart, A.~Abidin, S.~Gürses, M.~Veale, C.~Cremers,
  M.~Backes, N.~O. Tippenhauer, R.~Binns, C.~Cattuto, A.~Barrat, D.~Fiore,
  M.~Barbosa, R.~Oliveira, and J.~Pereira.
\newblock \enquote{{Decentralized Privacy-Preserving Proximity Tracing
  (dp3t)}}.
\newblock Technical Report 2005.12273, arXiv, 2020.

\bibitem{van2020aerosol}
N.~Van~Doremalen, T.~Bushmaker, D.~H. Morris, M.~G. Holbrook, A.~Gamble, B.~N.
  Williamson, A.~Tamin, J.~L. Harcourt, N.~J. Thornburg, S.~I. Gerber, et~al.
\newblock \enquote{Aerosol and surface stability of SARS-CoV-2 as compared with
  SARS-CoV-1}.
\newblock {\em New England Journal of Medicine}, 382(16):1564--1567, 2020.

\bibitem{van08}
P.~Van~Mieghem, J.~Omic, and R.~Kooij.
\newblock \enquote{Virus spread in networks}.
\newblock {\em IEEE/ACM Transactions On Networking}, 17(1):1--14, 2008.

\bibitem{wang03}
Y.~Wang, D.~Chakrabarti, C.~Wang, and C.~Faloutsos.
\newblock \enquote{Epidemic spreading in real networks: An eigenvalue
  viewpoint}.
\newblock In {\em IEEE International Symposium on Reliable Distributed
  Systems}, pages 25--34, 2003.

\bibitem{dyoung2020}
D.~G. Young.
\newblock \enquote{How Far Can You Go?}, 2020.
\newblock \url{http://www.davidgyoungtech.com/2020/05/15/how-far-can-you-go},
  \text{Accessed on 18/07/2020}.

\bibitem{youssef11}
M.~Youssef and C.~Scoglio.
\newblock \enquote{An individual-based approach to SIR epidemics in contact
  networks}.
\newblock {\em Journal of theoretical biology}, 283(1):136--144, 2011.

\bibitem{china-nyt}
R.~Zhong.
\newblock \enquote{China's Virus Apps May Outlast the Outbreak, Stirring
  Privacy Fears}.
\newblock {\em New York Times}, May 2020.

\end{thebibliography}

\newpage

\section*{Author Biographies}

~
\begin{wrapfigure}{l}{0.3\textwidth}
\includegraphics[width=0.3\textwidth]{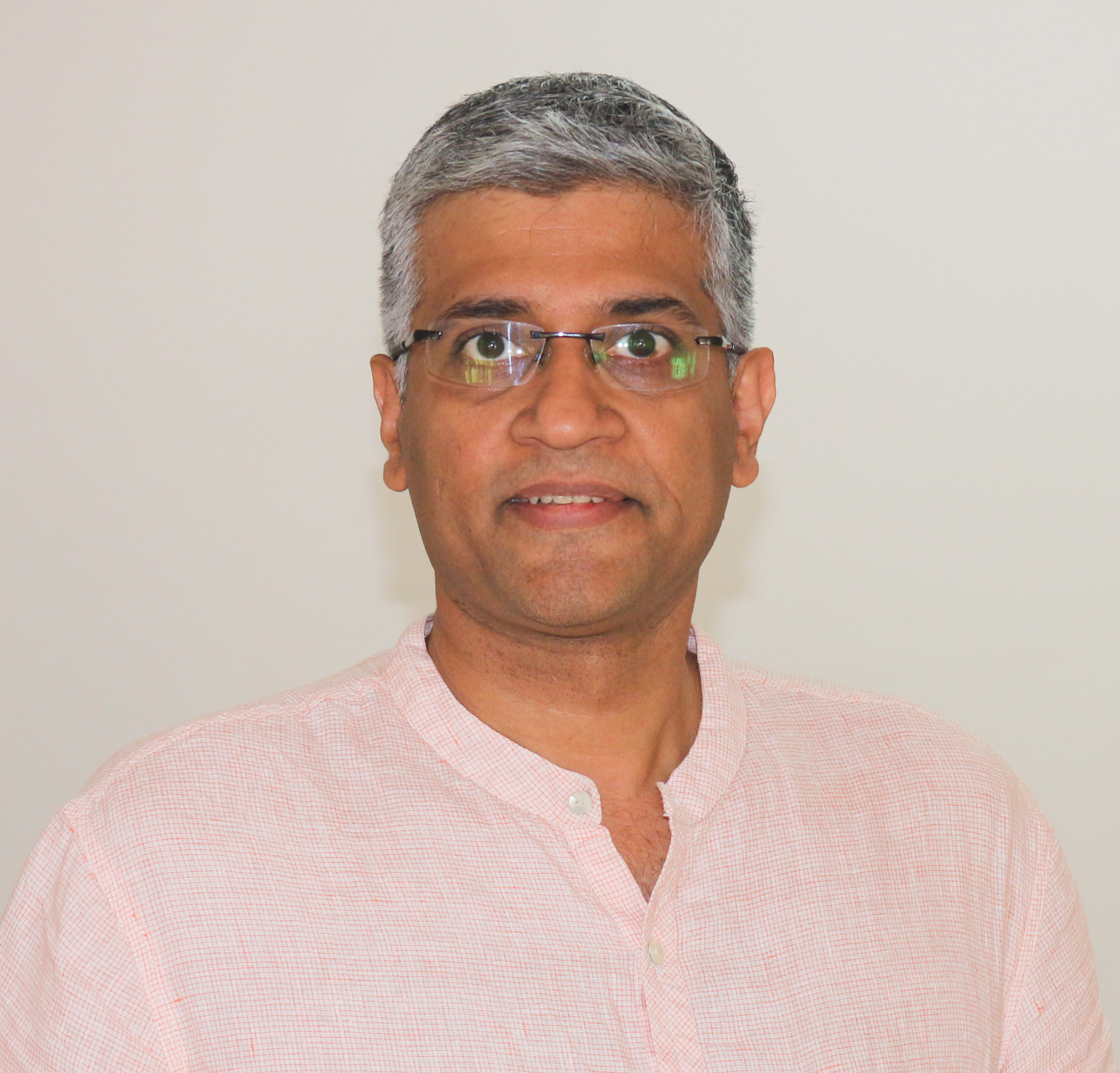}
\end{wrapfigure}
\noindent  \textbf{Yogesh Simmhan} is an Associate Professor in the Department of Computational and Data Sciences and a Swarna Jayanti Fellow at the Indian Institute of Science, Bangalore. His research explores abstractions, algorithms and applications on distributed systems. These span Cloud and Edge Computing, Graph Processing Platforms and Internet of Things (IoT) to support emerging data-driven applications. Yogesh has a Ph.D. in Computer Science from Indiana University, Bloomington, and was previously a Research Faculty at the University of Southern California (USC), Los Angeles, and a Postdoc at Microsoft Research, San Francisco. He is a Senior Member of the IEEE and the ACM.\\

\begin{wrapfigure}{l}{0.3\textwidth}
  \includegraphics[width=0.3\textwidth]{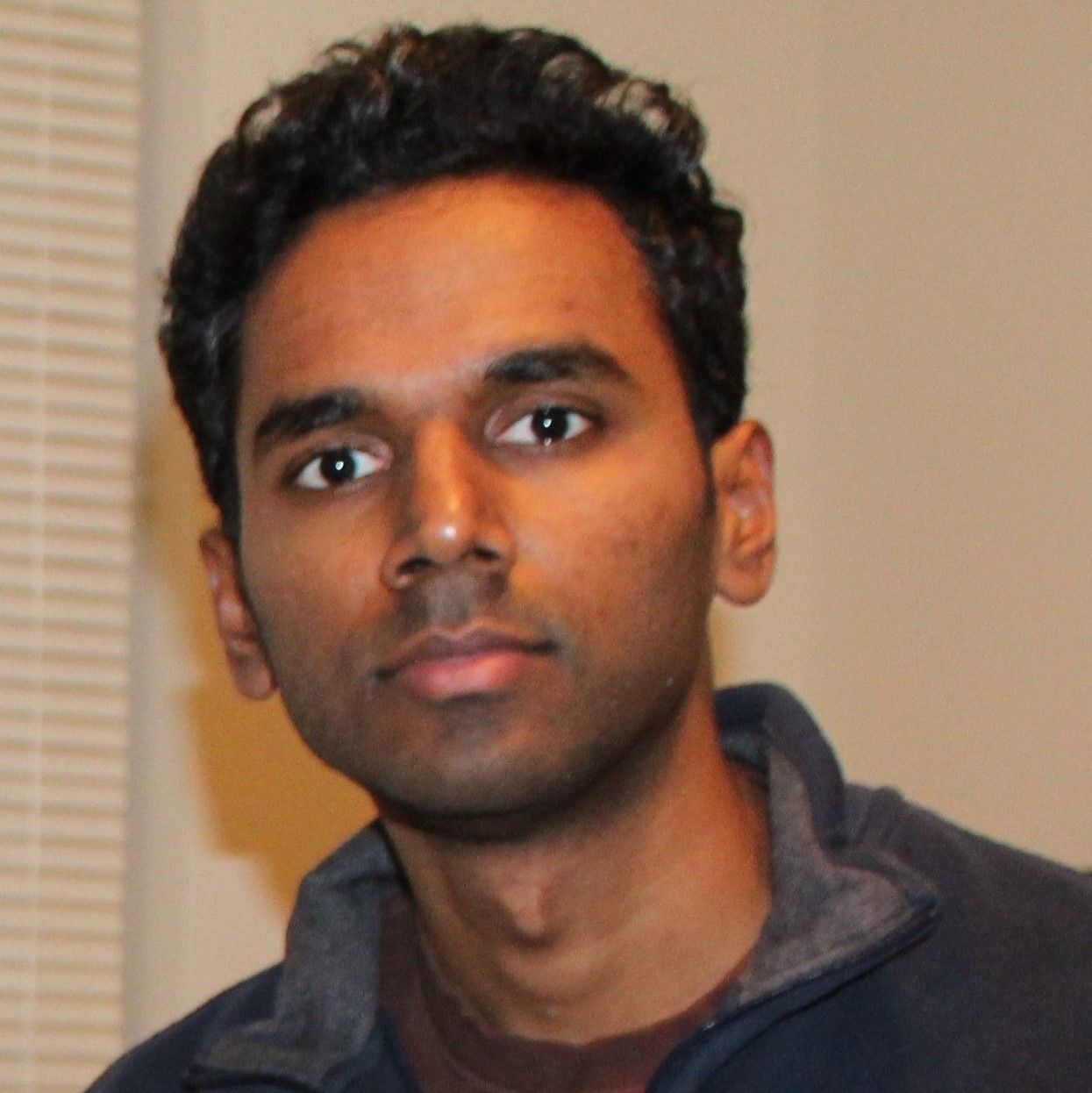}
\end{wrapfigure}
\noindent  \textbf{Tarun Rambha}
is an assistant professor in Civil Engineering at the Indian Institute of Science (IISc) Bangalore and an affiliate faculty at the Center for infrastructure, Transportation and Sustainable Urban Planning (CiSTUP). He received his PhD from the University of Texas at Austin where he worked on network equilibrium, congestion pricing, and adaptive routing in stochastic transit and traffic networks. Prior to joining IISc, he was a post-doctoral researcher at Cornell University where he studied optimizing hospital evacuations and predicting evacuation behaviour during hurricanes. His research interests include network science, optimization, and transportation.\\

\begin{wrapfigure}{l}{0.3\textwidth}
  \includegraphics[width=0.3\textwidth]{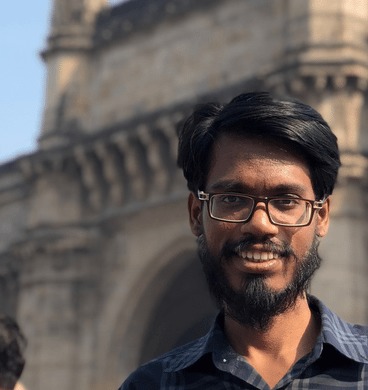}
\end{wrapfigure}
\noindent \textbf{Aakash Khochare} is a Ph.D. candidate at the Indian Institute of Science, Bangalore. His research interests are in distributed video analytics on edge and mobile computing platforms.\\
~\\
~\\
~\\
~\\
~\\
~\\
~\\

\begin{wrapfigure}{l}{0.3\textwidth}
  \includegraphics[width=0.3\textwidth]{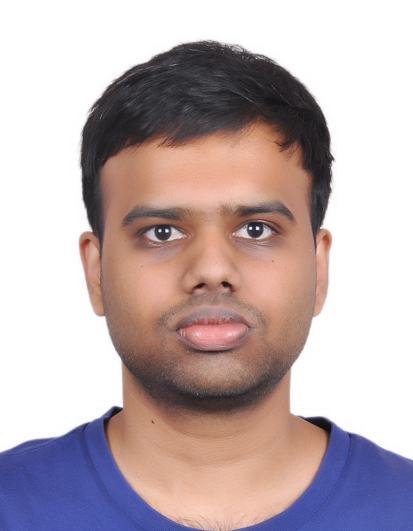}
\end{wrapfigure}
\noindent  \textbf{Shriram Ramesh} is a Graduate Student at the Department of Computational and Data Sciences at the Indian Institute of Science, Bangalore. He is supported by the Maersk CDS M.Tech. Fellowship. His research interests include graph processing, distributed systems and database systems. He was part of the team that won IEEE TCSC SCALE Challenge Award in 2019. He has a Bachelor's Degree in Electrical and Electronics Engineering and two years of consulting experience in the domain of Business Intelligence and Analytics.\\

\begin{wrapfigure}{l}{0.3\textwidth}
  \includegraphics[width=0.3\textwidth]{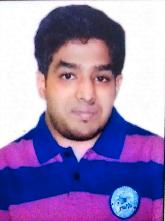}
\end{wrapfigure}
\noindent \textbf{Animesh Baranawal} is a Graduate Research Student in the Department of Computational and Data Sciences at the Indian Institute of Science, Bangalore. His research interests include graph processing and distributed systems. He has a Bachelor's Degree in Computer Science and Engineering and two years of industrial experience in Software Development.\\
~\\
~\\
~\\
~\\
~\\

\begin{wrapfigure}{l}{0.3\textwidth}
  \includegraphics[width=0.3\textwidth]{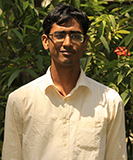}
\end{wrapfigure}
\noindent  \textbf{John V George} is a Junior Research Fellow working at the Center for infrastructure, Transportation and Sustainable Urban Planning (CiSTUP). He received his Bachelor's Degree in Computer Science and Engineering from NIT Calicut. His research interests are Machine Learning and Artificial Intelligence.\\
~\\
~\\
~\\
~\\
~\\
~\\
~\\
~\\
~\\
~\\
~\\
~\\

\begin{wrapfigure}{l}{0.3\textwidth}
  \includegraphics[width=0.3\textwidth]{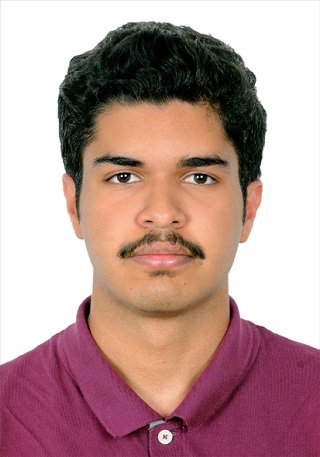}
\end{wrapfigure}
\noindent  \textbf{Rahul Atul Bhope} is a Project Associate at the DREAM Lab in the Department of Computational and Data Sciences at the Indian Institute of Science, Bangalore. He has a Bachelor's Degree in Instrumentation and Control Engineering from NIT Tiruchirappalli. His research interests include Internet of Things and Machine Learning with applications in Localization, Tracking and Healthcare.\\
~\\
~\\
~\\
~\\

\begin{wrapfigure}{l}{0.3\textwidth}
 \includegraphics[width=0.3\textwidth]{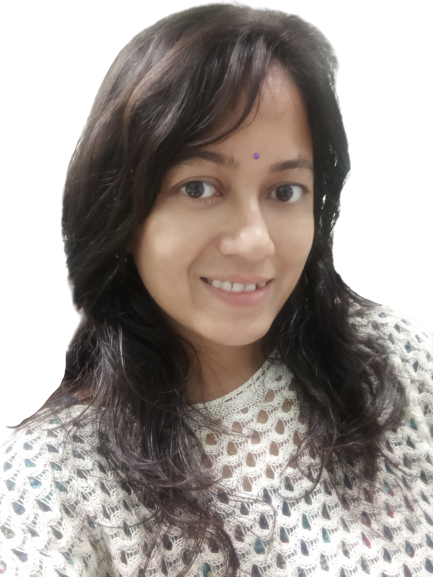}
\end{wrapfigure}
\noindent  \textbf{Amrita Namtirtha}  is a Research Associate at the DREAM Lab in the Department of 
Computational and Data Sciences at the Indian Institute of Science, Bangalore. She has a Ph.D. Degree in the Department of Computer Science and Engineering from the National Institute of Technology,
Durgapur. She worked in  TATA Consultancy Services Pvt. Ltd for about three years as a consultant of ERPLab and Oracle fusion. Her research interests include
social network analysis, graph theory, big data analysis, and image processing. She is a reviewer of reputed journals such as
Elsevier Physica A, IEEE Access, IEEE conferences.\\

\begin{wrapfigure}{l}{0.3\textwidth}
  \includegraphics[width=0.3\textwidth]{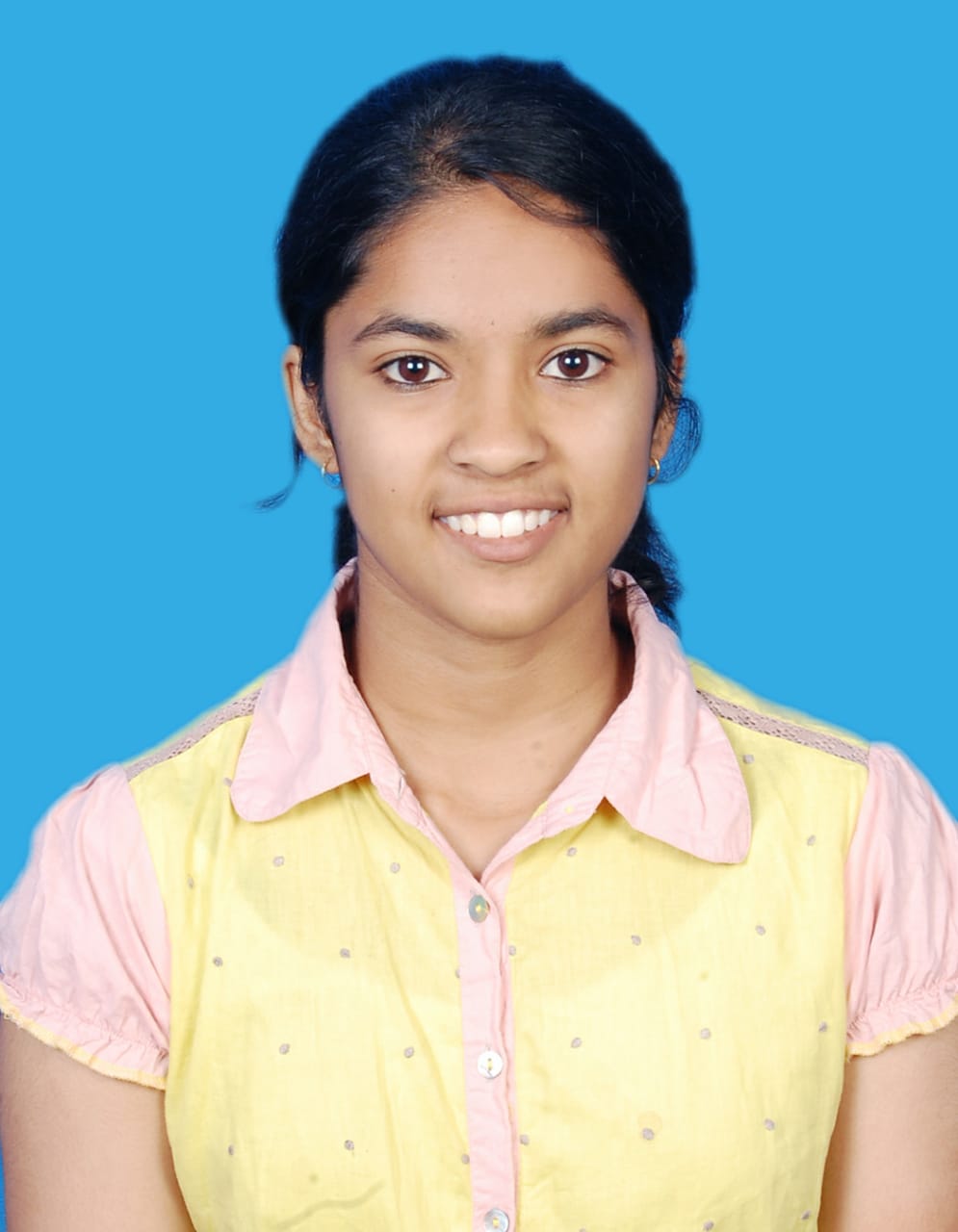}
\end{wrapfigure}
\noindent  \textbf{Amritha Sundararajan} is pursuing an integrated M.Sc in Theoretical Computer Science at PSG College of Technology, Coimbatore. Her research interests include graph processing and algorithms.

\clearpage

\setglossarystyle{altlist}
\printglossary[title=Special Terms, toctitle=List of terms]

\end{document}